\documentclass[aps, amsmath, amssymb, preprintnumbers, twocolumn ,noshowpacs, superscriptaddress]{revtex4-1}
\usepackage[USenglish]{babel}
\usepackage{xcolor}
\usepackage{array}
\usepackage{amsmath,amssymb,amsfonts,mathrsfs}
\usepackage{bm}
\usepackage{epsfig}
\usepackage[colorlinks=true,linkcolor=blue,citecolor=red]{hyperref}
\newcommand{\be}{\begin{equation}}
\newcommand{\ee}{\end{equation}}
\newcommand{\beS}{\begin{equation*}}
\newcommand{\eeS}{\end{equation*}}
\newcommand{\bea}{\begin{eqnarray}}
\newcommand{\eea}{\end{eqnarray}}
\newcommand{\ba}{\begin{eqnarray*}}
\newcommand{\ea}{\end{eqnarray*}}
\newenvironment{eqs}
{\begin{equation} \begin{aligned}}
{\end{aligned} \end{equation} }
\newcommand{\bal}{\begin{eqs}}
\newcommand{\eal}{\end{eqs}}

\newcommand{\bas}{\begin{eqs}}
\newcommand{\eas}{\end{eqs}}
\newcommand{\dagga}{{\phantom{\dagger}}}

\newcommand{\bQ}{\mathbf{Q}}

\newcommand{\bK}{\mathbf{K}}
\newcommand{\bq}{\mathbf{q}}

\newcommand{\bk}{\mathbf{k}}

\newcommand{\bp}{\mathbf{p}}

\newcommand{\bx}{\mathbf{x}}

\newcommand{\br}{\mathbf{r}}

\newcommand{\eqn}[1]{(\ref{#1})}
\newcommand{\ket}[1]{\mid\! #1\rangle}
\newcommand{\bra}[1]{\langle #1\!\mid} 
\newcommand{\bw}{\begin{widetext}}
\newcommand{\ew}{\end{widetext}}

\newcommand{\ep}{{\epsilon}}

\begin{document}
\title{Moir\'e lattice effects on the orbital magnetic response of twisted bilayer graphene and Condon instability}
\author{Daniele Guerci}
\affiliation{Universit{\'e} de Paris, Laboratoire Mat{\'e}riaux et Ph{\'e}nom{\`e}nes Quantiques, CNRS, F-75013 Paris, France.} 
\affiliation{Universit{\'e} Paris-Saclay, CNRS, Laboratoire de Physique des Solides, 91405, Orsay, France.} 
\author{Pascal Simon}
\affiliation{Universit{\'e} Paris-Saclay, CNRS, Laboratoire de Physique des Solides, 91405, Orsay, France.} 
\author{Christophe Mora}
\affiliation{Universit{\'e} de Paris, Laboratoire Mat{\'e}riaux et Ph{\'e}nom{\`e}nes Quantiques, CNRS, F-75013 Paris, France.} 
\affiliation{Dahlem Center for Complex Quantum Systems and Fachbereich Physik, Freie Universit\"at Berlin, 14195, Berlin, Germany}

\begin{abstract}

We analyze the orbital magnetic susceptibility from the band structure of twisted bilayer graphene. Close to charge neutrality, the out-of-plane susceptibility inherits the strong diamagnetic response from graphene. Increasing the doping, a crossover from diamagnetism to paramagnetism is obtained and a logarithmic divergence develops at the van Hove singularity of the Moir\'e lattice in the first band. The enhanced paramagnetism at the van Hove singularity is stronger for relatively large angle but gets suppressed by the flat spectrum towards the vicinity of the first magic angle. A diverging paramagnetic susceptibility indicates an instability towards orbital ferromagnetism with an orbital out-of-plane magnetization and a Landau level structure. The region of instability is however found to be practically very small, parametrically suppressed by the ratio of the electron velocity to the speed of light. 
We also discuss the in-plane orbital susceptibility at charge neutrality where we find a paramagnetic response and a logarithmic divergence at the magic angle. The paramagnetic response is associated with negative counterflow current in the two layers and does not admit a semiclassical description. Results at finite doping shows a logarithmic divergence of the in-plane orbital susceptibility at the van Hove singularity. Interestingly, in this case the paramagnetism is enhanced by approaching the magic-angle region.

%For an out-of-plane magnetic field we find  a large diamagnetic response close to charge neutrality, which resembles the one observed in graphene. 
%As we increase the doping we  predict a crossover from diamagnetism to paramagnetism. 
%Then, when the chemical potential is tuned to the Van Hove singularity of the Moir\'e lattice we find a paramagnetic logarithmic singularity. 
%In this regime the TBG is unstable towards the formation of Condon domains with orbital magnetization oriented perpendicularly to the quasi-2D material. 
%However, such an instability 
%By comparing with real numbers we find that the effect 
%turns out to be exponentially suppressed in the ratio $c/v_0$, where $v_0$ is the electronic velocity in graphene and $c$ the speed of light, and is therefore very small.
%and probably beyond experimental reach.
%Investigating the properties of the orbital magnetic susceptibility close to the Van Hove singularity we show that orbital paramagnetism to an out-of-plane magnetic field is suppressed in the flat band regime. 
%On the other hand, we find at charge neutrality a paramagnetic response to in-plane magnetic field, which diverges logarithmically at the magic angles. 
%Interestingly, the effect is a ground state property determined by the whole spectrum of TBG and cannot be predicted by a low-energy description.
\end{abstract}

\date{\today}

\maketitle

\section{Introduction}
\label{intro}

Twisted bilayer graphene (TBG) has emerged at the forefront of condensed matter physics. 
Overlaying two layers of graphene with a small relative angle results in a Moir\'e interference pattern that drastically changes the behavior of electrons, slowing them down and enabling them to interact in ways that change the materials electronic properties. 
Interestingly, for specific twist-angles, named magic angles, the low-energy bands become extremely flat and strong many-body effects lead to correlated insulating states and superconductivity \cite{Cao_2018,Cao_2018_ins}. 
The easily accessible control of the carrier concentration, of the flatness of the bands and the opportunity to stack different materials make this system a versatile platform for exploring unconventional phenomena. 

In addition to strongly correlated phases TBG has been predicted to host fragile topology \cite{Song_2019,Vishwanath_2019_fragile_top}, experimentally evidenced by topologically protected anomalous Hofstadter spectra at strong magnetic field \cite{Lu2020_multiple,Herzog_2020,Lian_2020_LLs}. Quite generally, this highlights the richness of physical phenomena uncovered by coupling the orbital motion of electrons in TBG to an external electromagnetic field \cite{Lu_2019,XiDai_PRX_2019,Li_PRB_2020}. 
%This theoretical prediction has been recently proved experimentally by studying the Hofstadter spectrum realized by applying a strong magnetic field perpendicular to the bilayer plane \cite{Lu2020_multiple}. 
TBG is an exotic material with purely orbital-based magnetism and correlated orbital ferromagnetic states \cite{wu2020,Polshyn_2020,tschirhart2020imaging}. The existence and competition between different ferromagnetic and Chern insulating phases depends crucially on the alignment to the substrate \cite{Sharpe_2019}, the presence of a magnetic field or strain \cite{zondiner2020,wong2020,saito2021,choi2020}. Intrinsic Chern orbital phases have also be observed in TBG \cite{Serlin_2020,nuckolls2020,stepanov2020}. The nature of the ground state emerges from an intricate competition governed by Coulomb interaction, kinetic energy and topology, with many symmetry-breaking states of nearby energy \cite{kang2018,ahn2019,kang2019,Seo2019,repellin2020,Zalatel_2020,pons2020,zhang2020,liu2021theories,MacDonald_2020,bultinck2020,liu2020nematic}.

%This remarkable discovery highlights the richness of physical phenomena that one uncovers by coupling the orbital motion of electrons in TBG to %an external electromagnetic field \cite{Herzog_2020,Lian_2020_LLs}.  
%Moreover, TBG is to our knowledge the first example of a material with purely orbital-based magnetism, that gives rise to a correlated orbital %ferromagnetic insulator \cite{,Lu_2019,Serlin_2020,saito2021,nuckolls2020,wu2020,Polshyn_2020,tschirhart2020imaging}.
%The nature of the insulating phase is determined either by the combined effect of the substrate, which breaks the $\mathcal{C}_{2z}$ symmetry, %and of the electron-electron interaction which favours one of the two valleys \cite{XiDai_PRX_2019,Li_PRB_2020,Zalatel_2020,liu2021theories} %or by electronic correlations only \cite{MacDonald_2020,liu2020nematic}. 

In this paper, we investigate the orbital magnetic susceptibility in TBG as obtained from the band spectrum. The motivation also comes from a recent experiment \cite{vallejo2020detection} where the singular diamagnetism of monolayer graphene has been measured using a giant magnetoresistance sensor. In addition to the orbital magnetic response to an out-of-plane magnetic field $\chi^{\perp}_{\text{OMS}}$, TBG shows a finite orbital magnetic susceptibility to an in-plane magnetic field $\chi^{\parallel}_{\text{OMS}}$ that originates from the interlayer motion of the electrons.
Using the continuum model derived in Refs. \cite{Castro_Neto_2007,Morell_2010,Bistritzer_2011} and linear response theory, we compute $\chi^{\parallel}_{\text{OMS}}$ and $\chi^{\perp}_{\text{OMS}}$ as a function of the twisting angle and electron doping. The out-of-plane susceptibility $\chi^{\perp}_{\text{OMS}}$ is diamagnetic in the vicinity of charge neutrality but crossovers to a paramagnetic response upon electron (or hole) doping and exhibits a logarithmic singularity when the Fermi energy is tuned to the van Hove singularity (VHS) of the lowest conduction band. As pointed out by Vignale \cite{Vignale_1991}, the paramagnetic behavior originates from the saddle-point structure close to the VHS, where the quasiclassical orbits circle anti-clockwise and enhance the applied field. The properties of the VHS depend on the angle of twisting and a higher-order VHS \cite{Fu_2019} occurs around $1.11^{\circ}$, slightly above the first magic angle. We find that the resulting band flattening at the higher-order VHS is detrimental to the paramagnetic singular behavior and the divergence is gradually smeared out when decreasing the twisting angle towards $1.11^{\circ}$. 

Interestingly, a very strong paramagnetic response is associated with a ground state instability towards the formation of magnetic domains, called Condon domains \cite{Condon_1966,Azbel_1968,Holstein_1973,Markiewicz_1985,Quinn1985,Wyder_1998}, as recently discussed in Ref. \cite{Andolina_2020,Basko_PRL2019}. It implies to take into account the usually small "Amperean" magnetic interaction \cite{kargarian2016} which occurs between charged electric currents \cite{Jaksch_2019} and can be enhanced by embedding the system into a cavity. 
We revisit this problem by treating the interaction between the cavity photons and the orbital currents within the random phase approximation and recover precisely the onset of Condon phase instability \cite{Andolina_2020} for the out-of-plane component of the magnetic field. We also introduce a criterion for the occurrence of photon condensation for a magnetic field lying in the plane of the top and bottom layers. Proceeding further with a mean-field analysis, we obtain a time-reversal symmetry-broken phase with an effective magnetic field opening Landau levels and a finite Chern number. It exhibits a non-zero orbital magnetization which generates the magnetic field.
 With respect to standard electron-electron interaction, the Amperean energy is small by a factor corresponding to the ratio between the typical electron velocity and the speed of light squared \cite{Basko_PRL2019}. As a result, the Condon instability is restricted to an extremely small region. The same small parameter explains the removal of the logarithmic paramagnetic singularity at the higher-order VHS since flat bands entail small electron velocities.

%Moreover, we show by performing analytical calculations that $\chi^{\parallel}_{\text{OMS}}$ is paramagnetic at charge neutrality and it is enhanced close to the magic angles. 
%On the other hand, $\chi^{\perp}_{\text{OMS}}$ shows a paramagnetic logarithmic singularity when the Fermi energy is tuned with the Van Hove singularity (VHS) of the lowest energy band in TBG. 
%In the presence of such a large magnetic response it is important to take into account the âAmperean" magnetic interaction which occurs between charged electric currents. 
%By treating such an interaction within the random phase approximation we derive a simple criterion for the formation of an orbital ferromagnetic ground state which coincides with the inequality for the Condon domain phase \cite{Condon_1966,Azbel_1968,Holstein_1973,Wyder_1998}. 
%Interestingly, as a result of the logarithmic singularity in the orbital magnetic susceptibility at the VHS the TBG is unstable towards the formation of magnetic domains characterized by a finite out-of-plane orbital magnetization. 

Then, we examine the in-plane susceptibility $\chi^{\parallel}_{\text{OMS}}$, primarily at charge neutrality. In contrast with the out-of-plane component, it exhibits a paramagnetic response and it is enhanced close to magic angles, in agreement with Ref. \cite{Stauber_2018}. We analytically derive a logarithmic divergence at the magic angle. From Maxwell's equations, one can relate $\chi^{\parallel}_{\text{OMS}}$ to the Drude weight of the counterflow conductivity, corresponding to the current imbalance between the two layers as a result of a perpendicular electric field gradient. A paramagnetic behaviour then corresponds to a negative Drude weight eluding a semi-classical description. Away from charge neutrality we find a logarithmic singularity of $\chi^{\parallel}_{\text{OMS}}$ which becomes a power-law divergence at the higher-order VHS.

%Remarkably and as a result of a coherent drag, the counterflow conductivity
%is negative corresponding to a paramagnetic behavior.

The paper is organized as follows: In Sec. \ref{OMS_quasi2D} we introduce the orbital magnetic response by considering quasi-2D materials. 
This is followed by the introduction to the continuum model employed to describe the electronic properties of TBG in Sec. \ref{continuum_TBG}. 
The coupling to the external magnetic field is discussed in Sec. \ref{coupling_external_field}, we also present the light-matter interaction obtained in a different gauge in Appendix \ref{in_plane_Aperp_gauge}. 
We then present our results for the orbital magnetic response to an out-of-plane magnetic field in Sec. \ref{OMS_out_of_plane}. 
The criterion for the instability towards the formation of an orbital ferromagnetic ground state is given in Sec. \ref{Condon_domains_TBG}.
Sec. \ref{OMS_in_plane} is devoted to the response of TBG to an in-plane magnetic field. 
Appendices \ref{interband_q_expansion}, \ref{intraband_q_expansion}, \ref{finite_size_interlayer} and \ref{reg_interlayer} contain useful details for the evaluation of the orbital magnetic susceptibilitites while in Appendix \ref{cavity_MODES} we detail the derivation of the criterion for the occurrence of photon condensation.
Finally, we conclude the discussion in Sec. \ref{Con} by briefly summarizing the scope of the work and setting the context for future work.

\section{The orbital magnetic response of quasi-2D materials}
 \label{OMS_quasi2D}

Twisted quasi-2D materials, stretched along the $(x,y)$ plane, are characterised due to the presence of interlayer hopping processes by an in-plane orbital magnetic response, in addition to the conventional out-of-plane one which has been extensively studied in various two-dimensional systems \cite{Stauber_2011,Raoux_2014,Raoux_2015,Yang_2015,Piechon_2016,Guinea_2016,PhysRevB.103.195104}. 
\begin{figure}
\begin{center}
\includegraphics[width=0.45\textwidth]{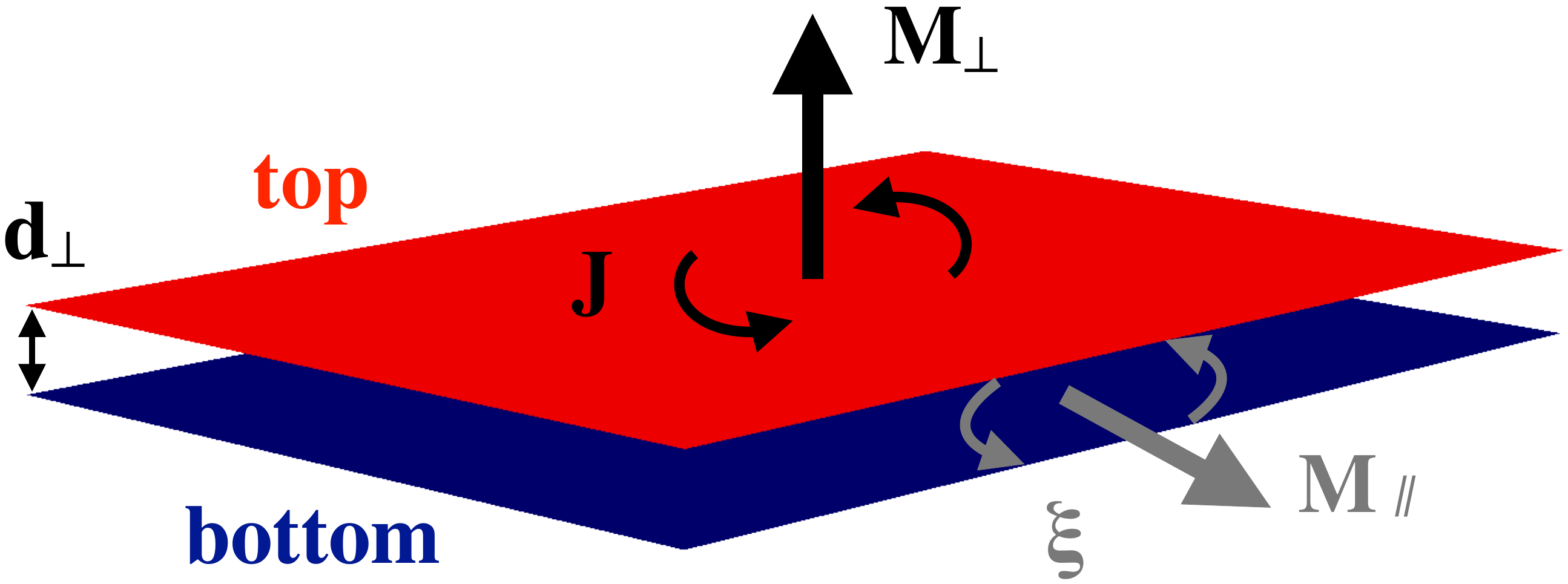}
\caption{Orbital magnetization response of a quasi-2D material. 
Red and blue planes refer to the bottom and top layer separated by an interlayer distance $d_\perp$. The in-plane solenoidal total charge current $\bm{J}$ gives rise to an out-of-plane magnetization $\bm{M}^{\text{orb}}_\perp$. On the other hand the solenoidal dipolar current $\bm{\xi}$ induces an in-plane magnetization $\bm{M}^{\text{orb}}_\parallel$.}
\label{sketch}
\end{center}
\end{figure}
In Fig. \ref{sketch} we describe schematically the orbital magnetizations and the corresponding current densities.   
The out-of-plane magnetization $\bm{M}^{\text{orb}}_\perp$ is related to the in-plane total charge current $\bm{J}$, while the in-plane one $\bm{M}^{\text{orb}}_\parallel$ to the dipolar current $\bm{\xi}$ that is defined as the current unbalance between top and bottom layers.
Within linear response theory, the orbital magnetic response tensor to an uniform magnetic field reads:
\be
\label{OMS_linear_response}
\left(\begin{array}{c}M^{\text{orb}}_x \\ M^{\text{orb}}_y \\ M^{\text{orb}}_z \end{array}\right)= \left(\begin{array}{ccc} \chi^\parallel_{\text{OMS}} & 0 & 0 \\0 & \chi^\parallel_{\text{OMS}} & 0 \\0 & 0 &  \chi^\perp_{\text{OMS}} \end{array}\right)\,\left(\begin{array}{c}B_x \\ B_y \\ B_z \end{array}\right),
\ee
where $\chi^\parallel_{\text{OMS}}$ and $\chi^\perp_{\text{OMS}}$ are the orbital magnetic susceptibilities (OMSs) to an in-plane and an out-of-plane magnetic fields, respectively. 
The orbital magnetic susceptibility tensor \eqn{OMS_linear_response} in the $\mathbf{q}\to0$ limit possesses full rotational symmetry which implies that only diagonal terms are not vanishing. Moreover, due to the in-plane isotropy we have $\chi^{\text{OMS}}_{xx}(\mathbf{q}\to0)=\chi^{\text{OMS}}_{yy}(\mathbf{q}\to0)=\chi^{\text{OMS}}_{\parallel}(\mathbf{q}\to0)$ while $\chi^{\text{OMS}}_{zz}(\mathbf{q}\to0)=\chi^{\text{OMS}}_{\perp}(\mathbf{q}\to0)\neq\chi^{\text{OMS}}_{\parallel}(\mathbf{q}\to0)$. 
Finally, we observe that in the case of TBG the OMSs depend parametrically on the twist-angle $\theta$, satisfying the following parity constraints: $\chi^\parallel_{\text{OMS}}(-\theta)=\chi^\parallel_{\text{OMS}}(\theta)$ and $\chi^\perp_{\text{OMS}}(-\theta)=\chi^\perp_{\text{OMS}}(\theta)$.

The susceptibilities $\chi^\parallel_{\text{OMS}}$ and $\chi^\perp_{\text{OMS}}$ in Eq. \eqn{OMS_linear_response} are obtained by employing the linear response relation between the current density $\mathbf{j}$ and the vector potential $\mathbf{A}$ which gives rise to the magnetic field $\mathbf{B}$ through the equation $\mathbf{B}=\nabla\times\mathbf{A}$.
For a finite thin sheet of material the current response to a static but spatially modulated vector potential $\mathbf{A}(\br,z)$ reads:
\be
\label{current_response_finite_slab}
j_i(\bq,z)=e\int\,dz^\prime\,\,Q^{i,j}(\bq,z,z^\prime,\omega=0)\,A_j(\bq,z^\prime),
\ee
where $\bq=(q_x,q_y)$ is the in-plane wavevector, $Q^{i,j}(\bq,z,z^\prime,\omega=0)$ is the static limit of the current response tensor and the $z$ integral is extended over the width $d_\perp$ of the quasi-2D material. 
Following Ref. \cite{DTSon_2020} we define the 2D charge $J_i(\bq)$ and dipolar $\xi_i(\bq)$ currents as: 
\be
\label{charge}
J_i(\bq)=\int\,dz\,j_i(\bq,z),
\ee
\be
\label{dipolar}
\xi_i(\bq)=\int\,dz\,z\,j_i(\bq,z).
\ee
By making use of the relations $\bm{j}=-\nabla\times\bm{m}^\text{orb}/e$ and $\bm{M}^{\text{orb}}=\int dz\,\bm{m}^\text{orb}(z)$ we find that the out-of-plane magnetization density $M^{\text{orb}}_z$ is expressed in terms of 2D charge current by $J_i=-\epsilon_{ijz}\partial_j\,M^{\text{orb}}_z/e$, while the in-plane magnetization $\bm{M}^{\text{orb}}_{\parallel}=(M^{\text{orb}}_x,M^{\text{orb}}_y)$ is related to the 2D dipolar current by $\bm{M}^{\text{orb}}_{\parallel}=e\,\pmb{\xi}\times\mathbf{z}$. 
Thanks to the previous relations and Eq. \eqn{current_response_finite_slab} we have:
\bal
\label{OMS_perp}
\chi^{\perp}_{\text{OMS}}&=-\frac{e^2}{2}\int dz\,\int dz^\prime\frac{\partial^2 }{\partial q^2}Q^{y,y}(q\,\mathbf{x},z,z^\prime,0)\Big|_{q=0},\\
&=-\frac{e^2}{2}\int dz\,\int dz^\prime\frac{\partial^2 }{\partial q^2}Q^{x,x}(q\,\mathbf{y},z,z^\prime,0)\Big|_{q=0},
\eal
and
\bal
\label{OMS_paral}
\chi^{\parallel}_{\text{OMS}}&=-e^2\int dz\, z\int dz^\prime\,z^\prime\,Q^{x,x}(\bq\to\bm{0},z,z^\prime,0),\\
&=-e^2\int dz\, z\int dz^\prime\,z^\prime\,Q^{y,y}(\bq\to\bm{0},z,z^\prime,0).
\eal
%The in-plane magnetization $\mathbf{M}^{\text{orb}}_{\parallel}$ corresponds to a finite dipolar current $\pmb{\xi}$ that describes currents flowing in opposite directions in the top and the bottom layers of the quasi-2D material. 
%As we will see in the following, the aforementioned correspondence introduces a relation between the susceptibility $\chi^{\parallel}_{\text{OMS}}$ and the counterflow conductivity $\sigma^{\text{CF}}$ introduced in Ref. \cite{Bistritzer_2011} and investigated in Refs. \cite{Stauber_2018,Stauber_2018_new}. 
We notice that Eqs. \eqn{OMS_perp} and \eqn{OMS_paral} are obtained by computing the response of the system to a long wavelength $\bq$ magnetic field $\mathbf{B}$ \cite{Vignale_1991,Mauri_1996,Niu_Vignale_2007}. 
Before analyzing the rich behaviour exhibited by the orbital magnetic susceptibilities in TBG, let us first detail the continuum model for TBG and the coupling to the magnetic field.    
  
\section{Continuum model for TBG}
\label{continuum_TBG}

In this work we describe TBG by employing the continuum Hamiltonian introduced in \cite{Castro_Neto_2007,Morell_2010,Bistritzer_2011} which consists of two layers of graphene described by Dirac fields at $\bK$ points of each layer and coupled through the Moir\'e potential $T(\br)$:
\be
\label{BM_TBG}
H_0 = \left(\begin{array}{cc} \hbar v_0\,\bm{\sigma}\cdot\hat{\mathbf{k}} & T(\br) \\ T^\dagger(\br) &  \hbar v_0\,\bm{\sigma}\cdot\hat{\mathbf{k}} \end{array}\right),
\ee
where $v_0$ is the bare Fermi velocity of monolayer graphene, $\hat{\mathbf{k}}=-i(\partial_x,\partial_y)$, $T(\br)=\sum_{j=1}^{3}\,T^j\,e^{i\bq_j\cdot\br}$, $\br$ is the electron coordinate in the 2D plane, $\bq_1=k_\theta(0,1)$, $\bq_{2,3}=k_\theta(\mp\sqrt{3}/2,-1/2)$ give rise to the Moir\'e lattice with modulation $k_\theta=2k_D\sin\theta/2$, $k_D=4\pi/3a$ is the Dirac momentum, with $a=2.46\text{\AA}$ being the lattice constant of monolayer graphene. In the previous expression we have introduced the matrices: 
\be
\label{T_matrices}
T^{j+1}=w\left[r\,\sigma^0+\cos\frac{2\pi j}{3}\,\sigma^x+\sin\frac{2\pi j}{3}\,\sigma^y\right],
\ee
where $w=110\, \text{meV}$ $(w\,r)$ parameterizes the $AB/BA$ $(AA/BB)$ interlayer hopping strengths, and $r<1$ due to lattice relaxation effects \cite{Koshino_2017,Koshino_2018}. Unless specified otherwise, the numerical evaluations of this paper have been performed with $r=0.7$.
The Hamiltonian originating from the graphene Dirac cones at $\bK^\prime$ is simply the time-reversal counterpart of $H_0$ \eqn{BM_TBG}. 
Within our description we neglect inter-valley hopping processes. 
\begin{figure}
\begin{center}
\includegraphics[width=0.45\textwidth]{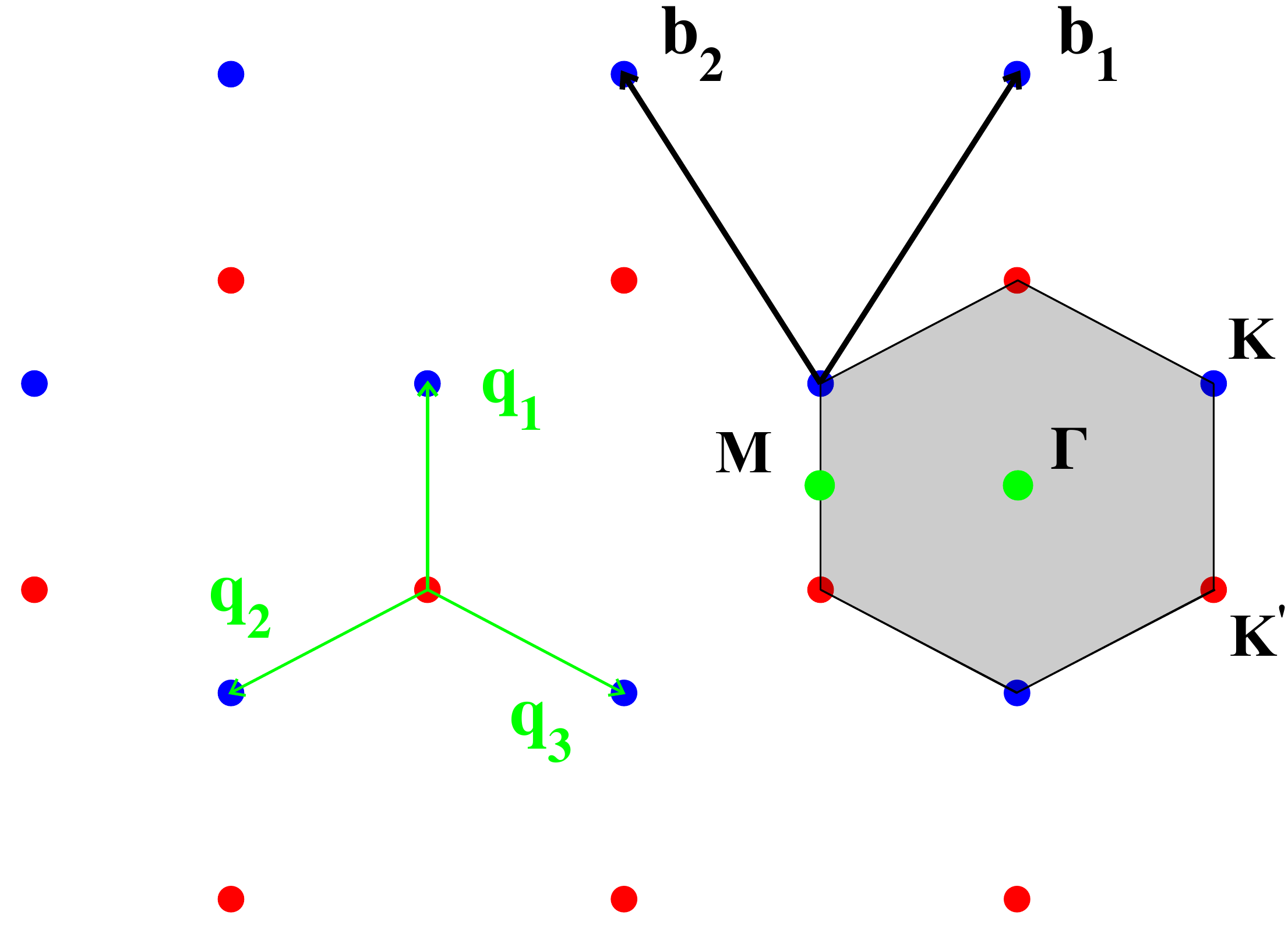}
\caption{The $\bk$-space lattice generated by $\bq_1$, $\bq_2$ and $\bq_3$. The honeycomb lattice is composed by two interpenetrating triangular sublattices $\mathcal{Q}_1$ and $\mathcal{Q}_2$ corresponding to bottom (blue dots) and top (red dots) layer, respectively. The grey shaded region centered at $\bm{\Gamma}$ depicts the MBZ, $\bK$ and $\bK^\prime$ are the high-symmetry points at the corner of the MBZ, while the $\mathbf{M}$ point is the middle point between $\bK$ and $\bK^\prime$. The vectors $\mathbf{b}_1=\bq_1-\bq_2$ and $\mathbf{b}_2=\bq_1-\bq_3$ are the primitive vectors of the triangular lattice.}
\label{lattice_TBG}
\end{center}
\end{figure}
In the second quantization formalism \cite{bernevig2020tbg,song2020tbg} the Hamiltonian \eqn{BM_TBG} reads: 
\be
\label{BM_TBG_2nd}
\hat{H}_0=\sum_{\bk\in\text{MBZ}}\sum_{\alpha\beta}\sum_{\bQ,\bQ^\prime\in\mathcal{Q}_{1/2}}c^\dagger_{\bk,\bQ,\alpha}\,H_{\bQ\alpha,\bQ^\prime\beta}(\bk)\,c^\dagga_{\bk,\bQ^\prime,\beta},
\ee
where $H_{\bQ\alpha,\bQ^\prime\beta}(\bk)$ is obtained by projecting $H_0$ in the plane wave basis: 
\bal
\label{BM_TBG_1st}
H_{\bQ\alpha,\bQ^\prime\beta}(\bk)=&\delta_{\bQ,\bQ^\prime}\,\hbar v_0\,(\bk-\bQ)\cdot\pmb{\sigma}_{\alpha\beta}\\
&+\sum_{j=1}^{3}\,T^j_{\alpha\beta}(\delta_{\bQ-\bQ^\prime,\bq_j}+\delta_{\bQ^\prime-\bQ,\bq_j}).
\eal
In the previous expressions \eqn{BM_TBG_2nd} and \eqn{BM_TBG_1st} $\alpha$ and $\beta$ denote the sublattice indices ($\alpha,\beta=A,B$), MBZ stands for the Moir\'e BZ (grey shaded region in Fig. \ref{lattice_TBG}), $l=1(2)$ denotes the bottom(top) layer, $\mathcal{Q}_{1/2}$ is the triangular lattice for layer $l=1/2$ depicted as blue and red dots in Fig. \ref{lattice_TBG}. 
Finally, we denote with $\ket{u_{n\bk}}$ and $\ep_{n\bk}$ the eigenstates and eigenvalues of $H_0$ \eqn{BM_TBG}. 

Notice that $H_0$ \eqn{BM_TBG} has only two dimensionless parameter $\alpha=w/\hbar v_0\,k_\theta$ and $r$.

 \section{Coupling to the magnetic field}
\label{coupling_external_field}

In the Coulomb gauge ($\nabla\cdot\mathbf{A}$=0) the effect of the magnetic field $\mathbf{B}=\nabla\times\mathbf{A}$ is introduced by the minimal substitution $\hat{\bk}\to\hat{\bk}+e\mathbf{A}(\br,z)/\hbar$ which gives the Hamiltonian: 
\be
H=H_0+\delta H,
\ee
with
\be
\delta H = e v_0 \left(\begin{array}{cc} \bm{\sigma}\cdot\mathbf{A}_1(\br) & 0 \\  0 &  \bm{\sigma}\cdot\mathbf{A}_2(\br) \end{array}\right),
\ee
and $\mathbf{A}_{1(2)}(\br)$ is the vector potential evaluate at the bottom (top) layer, i.e. $\mathbf{A}_{1(2)}(\br)=\mathbf{A}(\br,\mp d_\perp/2)$.  
Following Refs. \cite{Koshino_2017,Koshino_2018} we take into account the interlayer spacing $d_\perp$ constant ($d_\perp=3.433\text{\AA}$) and we account the effect of out-of-plane lattice distorsions by introducing the parameter $r$ in Eq. \eqn{T_matrices}.
In the second quantized formalism the Hamiltonian $\delta H$ is:
 \bal
 \label{delta_H}
 \delta \hat{H}=&e\sum_{\bq}\hat{\bm{J}}(\bq)\cdot\left[\mathbf{A}_1(-\bq)+\mathbf{A}_2(-\bq)\right]/2\\
 &+\frac{e}{d_\perp}\sum_{\bq}\hat{\pmb{\xi}}(\bq)\cdot\left[\mathbf{A}_2(-\bq)-\mathbf{A}_1(-\bq)\right],
 \eal
 where the charge \eqn{charge} and dipolar \eqn{dipolar} current operator are given by:
\bal
\label{charge_dipolar}
&\hat{\bm{J}}(\bq)=\hat{\bm{J}}_1(\bq)+\hat{\bm{J}}_2(\bq),\\
&\hat{\pmb{\xi}}(\bq)=d_\perp\,\left[\hat{\bm{J}}_2(\bq)-\hat{\bm{J}}_1(\bq)\right]/2.
\eal
In the previous expression we have introduced the current operator for layer $l$ that reads
\be
\label{current_layer}
\hat{J}_{i,l}(\bq)= v_0\,\sum_{\bk\in\text{MBZ}}\sum_{\bQ\in\mathcal{Q}_l}\sum_{\alpha,\beta}\,c^\dagger_{\bk,\bQ,\alpha}\,\sigma^i_{\alpha\beta}\,c^\dagga_{\bk+\bq,\bQ,\beta},
\ee 
where $i=x,y$ and $l=1(2)$ denotes the bottom(top) layer.
We notice that the dipolar current $\hat{\bm{\xi}}$ couples to the vector potential imbalance $\mathbf{A}_2-\mathbf{A}_1$ which gives rise to an in-plane magnetic field $B_i=\epsilon_{izj}(A_{j,2}-A_{j,1})/d_\perp$. 
Finally, by using the relation between the in-plane orbital magnetization $\bm{M}^\text{orb}_{\parallel}$ and $\bm{\xi}$ the second term can be written as a Zeeman interaction: 
 \bal
 \label{delta_H_1}
 \delta \hat{H}=&e\sum_{\bq}\hat{\bm{J}}(\bq)\cdot\left[\mathbf{A}_1(-\bq)+\mathbf{A}_2(-\bq)\right]/2\\
 &-\sum_{\bq}\hat{\bm{M}}^{\text{orb}}_\parallel(\bq)\cdot\mathbf{B}_\parallel(-\bq).
 \eal
 On the other hand, the charge current $\bm{J}$ is coupled to the total vector potential $\mathbf{A}_1+\mathbf{A}_2$.
 As explained in Appendix \ref{in_plane_Aperp_gauge}, a different choice of gauge is possible to discuss the orbital effect of the in-plane magnetic field.

In what follows we compute the orbital magnetic susceptibilities $\chi^\perp_{\text{OMS}}$ and $\chi^\parallel_{\text{OMS}}$. 

\section{The OMS to an out-of-plane magnetic field in TBG}
\label{OMS_out_of_plane}

\subsection{General expression and numerical results}

In terms of the charge current operator $\hat{\bm{J}}$ introduced in Eq. \eqn{charge_dipolar} the orbital magnetic susceptibility in Eq. \eqn{OMS_perp} takes the general form: 
\bal
\label{OMS_perp_1}
&\chi^{\perp}_{\text{OMS}}=-\frac{g_s\,g_v\,e^2}{2}\frac{\partial^2}{\partial q^2}\Bigg[\sum_{n,n^\prime}\int_{\text{MBZ}}\frac{d^2\bk}{4\pi^2}\\
&\frac{f(\ep_{n\bk})-f(\ep_{n^\prime\bk+q\bx})}{\ep_{n\bk}-\ep_{n^\prime\bk+q\bx}+i0^+}|\bra{u_{n\bk}}J^y\ket{u_{n^\prime\bk+q\bx}}|^2\Bigg]_{q=0},
\eal
where $J^y_{\bQ\alpha,\bQ^\prime\beta}=v_0\,\delta_{\bQ,\bQ^\prime}\,\sigma^y_{\alpha\beta}$, $f(\ep)=[e^{\beta(\ep-\mu)}+1]$ is the Fermi-Dirac distribution function and $g_v=g_s=2$ are the valley and spin degeneracies.
Following Ref. \cite{Vignale_1991} we write $\chi^{\perp}_{\text{OMS}}$ as a sum of the intraband ($n^\prime=n$) contribution and the interband ($n^\prime\neq n$) one: $\chi^\perp_{\text{OMS}}=\chi^{\perp,\text{intra}}_{\text{OMS}}+\chi^{\perp,\text{inter}}_{\text{OMS}}$. 
The intraband contribution $\chi^{\perp,\text{intra}}_{\text{OMS}}$ reduces to the following integral on the Fermi line:
\bal
\label{OMS_perp_intra}
\chi^{\perp,\text{intra}}_{\text{OMS}}=&-\frac{g_s\,g_v\,e^2}{4\pi^2\hbar^2}\sum_n\int _{\mathcal{S}_{n,\mu}}\frac{dl}{|\nabla_\bk\ep_{n\bk}|}\Bigg[\frac{\text{det}M^{-1}_{n\bk}}{12}\\
&-2\partial_{k_y}\ep_{n\bk}\,\lambda^{(1)}_n(\bk)-|\lambda^{(2)}_{n}(\bk)|^2\Bigg],
\eal
with $\ep_{n\bk}$ energy bands of the unperturbed Hamiltonian $H_0$ \eqn{BM_TBG}, $\mathcal{S}_{n,\mu}$ constant energy contour $\ep_{n\bk}=\mu$ in the $n$th band. 
Moreover, in Eq. \eqn{OMS_perp_intra} we have introduced the determinant of the inverse mass tensor:
\be
\label{inverse_mass_tensor}
M^{-1}_{n\bk}=\left(\begin{array}{cc}\partial^2_{k_x}\ep_{n\bk} & \partial_{k_x}\partial_{k_y}\ep_{n\bk} \\ \partial_{k_y}\partial_{k_x}\ep_{n\bk} & \partial^2_{k_y}\ep_{n\bk}\end{array}\right),
\ee  
while $\lambda^{(1)}_n(\bk)$ and $\lambda^{(2)}_n(\bk)$ are expressed in terms of the Bloch eigenstates and eigenvalues
\bal
\label{lambda_1}
\lambda^{(1)}_n(\bk)=&\bra{u_{n\bk}}\partial_{k_y}\,H_{0}\ket{u^{(2)}_{n\bk}}+\bra{u^{(2)}_{n\bk}}\partial_{k_y}\,H_0\ket{u_{n\bk}}\\
&-\bra{u^{(1)}_{n\bk}}\partial_{k_y}\,H_0+\partial_{y}\ep_{n\bk}\ket{u^{(1)}_{n\bk}},
\eal
\bal
\label{lambda_2}
\lambda^{(2)}_n(\bk)=&\bra{u_{n\bk}}\partial_{k_y}\,H_{0}\ket{u^{(1)}_{n\bk}}\\
&-\bra{u^{(1)}_{n\bk}}\partial_{k_y}\,H_{0}\ket{u_{n\bk}}.
\eal
In the Eqs. \eqn{lambda_1} and \eqn{lambda_2} we have introduced $\ket{u^{(1)}_{n\bk}}$ and $\ket{u^{(2)}_{n\bk}}$ that are defined by the perturbative expansion: 
\be
\ket{u_{n,\bk\pm q\,\mathbf{x}/2}}=\ket{u_{n\bk}}\pm q\,\ket{u^{(1)}_{n\bk}}+q^2\,\ket{u^{(2)}_{n\bk}}+\cdots.
\ee
Interestingly, it is possible to make progress and find an analytical expression for $M^{-1}_{n\bk}$, $\lambda^{(1)}_n(\bk)$ and $\lambda^{(2)}_n(\bk)$ in TBG as detailed in Appendix \ref{interband_q_expansion}.
Here we also prove that $\lambda^{(2)}_n(\bk)$ vanishes at all wavevectors as a consequence of the $\mathcal{C}_{2z}T$ symmetry of $H_0$ \eqn{BM_TBG}.   

On the other hand, the interband contribution is given by a sum over all occupied states: 
\bal
\label{OMS_perp_inter}
&\chi^{\perp,\text{inter}}_{\text{OMS}}=-\frac{g_s\,g_ve^2}{2}\Bigg[\frac{\partial^2}{\partial q^2}\sum_{n,n^\prime}^{n\neq n\prime}\int_{\text{MBZ}}\frac{d^2\bk}{4\pi^2}\\
&\frac{f(\ep_{n\bk})-f(\ep_{n^\prime\bk+q\bx})}{\ep_{n\bk}-\ep_{n^\prime\bk+q\bx}+i0^+}|\bra{u_{n\bk}}J^y\ket{u_{n^\prime\bk+q\bx}}|^2\Bigg]_{q=0}.
\eal
Analogously to the intraband term the second order derivative with respect to $q$ can be performed explicitly, details are left to Appendix \ref{interband_q_expansion}.
We stress that the evaluation of $\chi^{\perp,\text{inter}}_{\text{OMS}}$ is particularly challenging since the sums over the band indices $n$ and $n^\prime$ $(n\neq n^\prime)$ are unbounded. 
In order to perform the calculation we have introduced an UV cutoff that limits the sum over $n$ and $n^\prime$ to a finite number of bands. 
Then, we perform the finite size scaling to extrapolate the value of $\chi^{\perp,\text{inter}}_{\text{OMS}}$ by fitting the numerical data with the function $a+b/\sqrt{N}$ which has been derived from monolayer graphene.
Details on the finite size scaling analysis are deferred to Appendix \ref{finite_size_interlayer}.  

\begin{figure}
\begin{center}
\includegraphics[width=0.48\textwidth]{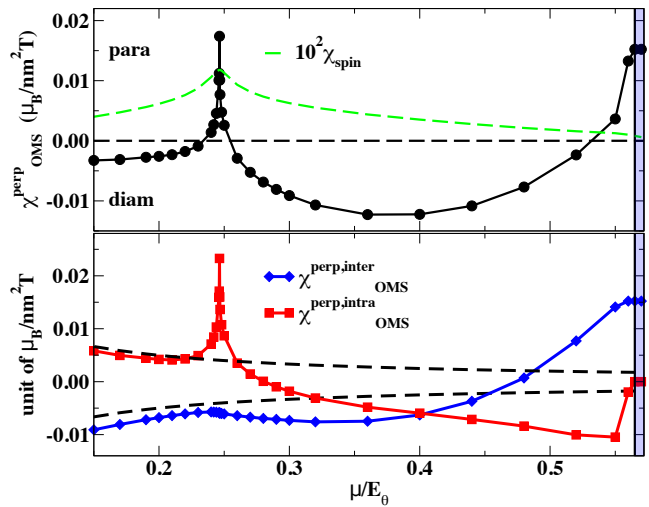}
\caption{Top: black data show the orbital magnetic susceptibility as a function of the Fermi energy $\mu/E_\theta$ ($E_\theta=\hbar v_0 k_\theta=484\,\text{meV}$) for $\theta=2.8^\circ$ $(\alpha\simeq0.227)$ and $r=0.7$. 
The black dashed line helps to distinguish the diamagnetic $\chi^\perp_{\text{OMS}}<0$ from the paramagnetic $\chi^\perp_{\text{OMS}}>0$ regime. 
Bottom: red and blue data correspond to the intraband and interband contributions to $\chi^\perp_{\text{OMS}}$, respectively. 
In this case the black dashed lines show the intraband and interband contributions for two decoupled Dirac cones, whose sum is exactly zero.
The blue dashed region denotes the band gap between the first and second positive energy band of TBG.
For comparison, we show in green the spin magnetic susceptibility $\chi_{\text{spin}}$ multiplied by $10^2$.
 We underline that the $\chi_{\text{spin}}$ is 2 order of magnitude smaller than $\chi^{\perp}_{\text{OMS}}$.
The value of $\chi^\perp_{\text{OMS}}$ is measured in unit of $\mu_B/\text{nm}^2\,\text{T}$, where $\mu_B$ is the Bohr magneton, $\text{T}$ is Tesla and refers to the intensity of the magnetic field.}
\label{perp_OMS_angle_28}
\end{center}
\end{figure}
\begin{figure}
\begin{center}
\includegraphics[width=0.45\textwidth]{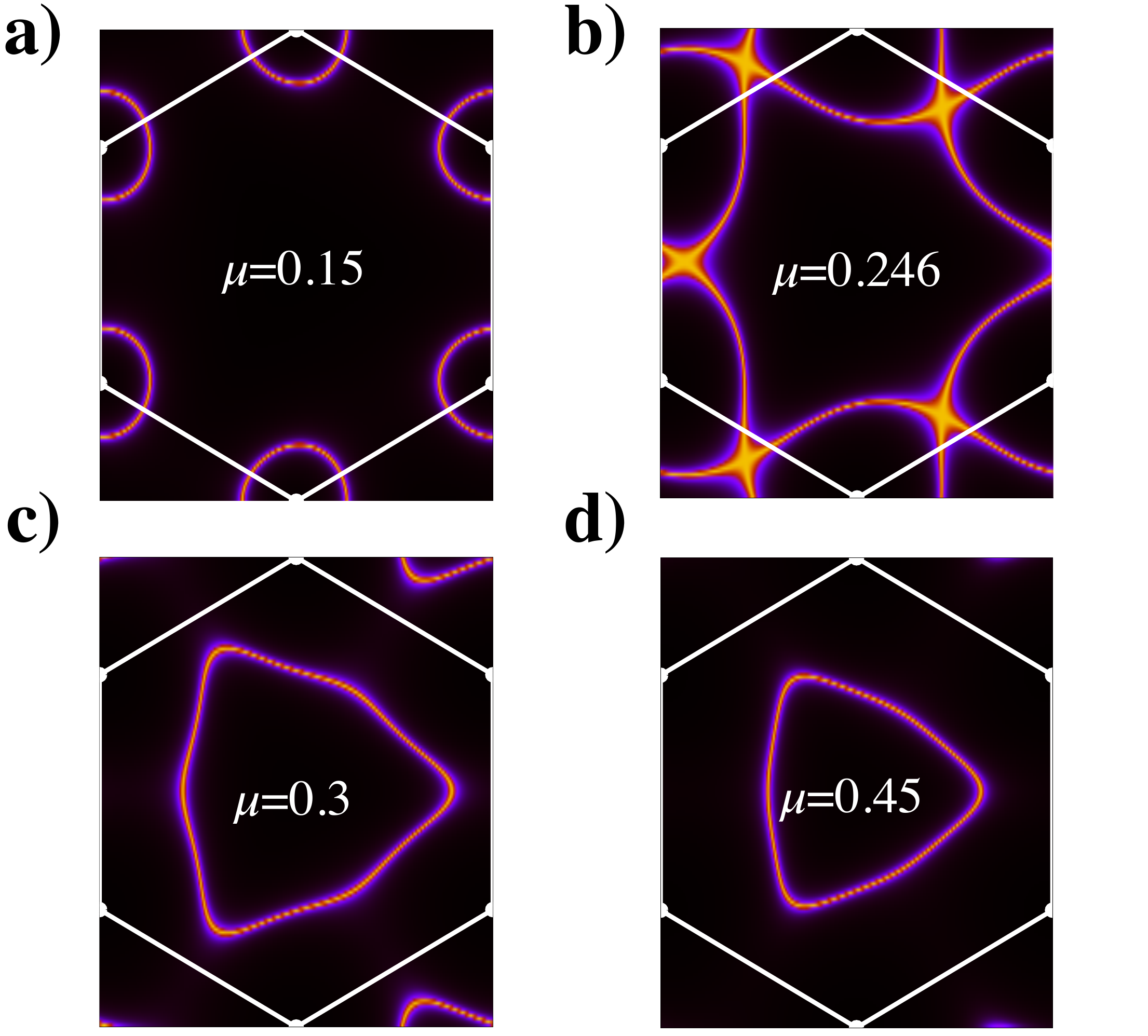}
\caption{Evolution of the Fermi surface within the lowest positive energy band $\ep_{+\bk}$ of TBG. In panels a), b) and c) we represent the Fermi surface for the values of the Fermi energy $\mu/\hbar v_0k_\theta=0.15, 0.246, 0.3$ and $0.45$. The white solid line denotes the boundary of the MBZ.}
\label{Fermi_surface_4}
\end{center}
\end{figure}
\begin{figure}
\begin{center}
\includegraphics[width=0.48\textwidth]{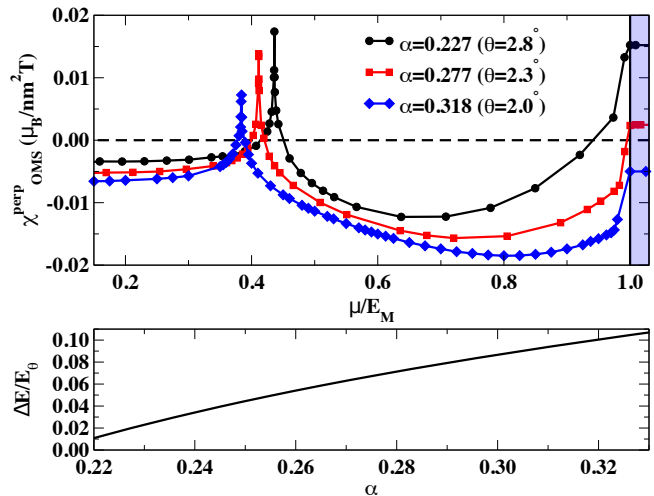}
\caption{Top: orbital magnetic susceptibility for different values of $\alpha$ as a function of $\mu/E_M$ where $E_M$ is the maximum of the lowest positive energy band $\ep_{+\bk}$. 
Bottom: energy gap between the first two positive energy bands $\Delta E$ in unit of $E_\theta=\hbar v_0 k_\theta$ as a function of $\alpha$. 
For small values of the energy gap $\Delta E$ we observe a crossover from diamagnetism to paramagnetism when approaching the top of the lowest positive energy band.
The value of $\chi^\perp_{\text{OMS}}$ is measured in unit of $\mu_B/\text{nm}^2\,\text{T}$, where $\mu_B$ is the Bohr magneton, $\text{T}$ is Tesla and refers to the intensity of the magnetic field.}
\label{perp_OMS_angle_alpha}
\end{center}
\end{figure}

The top panel in Fig. \ref{perp_OMS_angle_28} shows our result for the orbital magnetic susceptibility $\chi^\perp_{\text{OMS}}$ at twist-angle $\theta=2.8^\circ$ ($\alpha\simeq0.227$) as a function of the Fermi energy $\mu$ that varies within the lowest positive energy band of TBG for simplicity denoted as $\ep_{+\bk}$.   
In addition we show in the bottom panel of Fig. \ref{perp_OMS_angle_28} the intraband and interband contributions to $\chi^\perp_{\text{OMS}}$.
At charge neutrality TBG inherits the strong diamagnetism of graphene \cite{Koshino_2007,Koshino_2009,Principi_2009,Principi_2010} which is given by the contribution of the low-energy Dirac cones $\chi^\perp_{\text{OMS}}\simeq-g_s\,g_v e^2\,v^2\,\delta(\mu)/32\pi$, where $v$ is the electron velocity in TBG and $\delta(\mu)$ the Dirac delta function. As a consequence of the band flattening induced by the Moir\'e potential the prefactor of the delta function $\delta(\mu)$ is reduced by $v^2/v^2_0<1$ with respect to monolayer graphene.
Away from charge neutrality the Moir\'e potential removes the exact cancellation between $\chi^{\perp,\text{intra}}_{\text{OMS}}$ and $\chi^{\perp,\text{inter}}_{\text{OMS}}$ which occurs in monolayer graphene at finite doping \cite{Koshino_2007,Koshino_2009,Principi_2009,Principi_2010}. 
For small values of the chemical potential, Fig. \ref{Fermi_surface_4} a), the Fermi surface consists of two isolated circles located at the corners $\bK$ and $\bK^\prime$ of the MBZ in Fig. \ref{lattice_TBG}. 
Correspondingly in this regime, the interband term overcomes the intraband contribution and the overall response is diamagnetic, as shown in the top panel of Fig. \ref{perp_OMS_angle_28}. 
As the Fermi energy increases we observe a gradual crossover from diamagnetic to paramagnetism. 
The paramagnetic susceptibility diverges (logarithmically) at the van Hove singularity (VHS) which, for $\theta=2.8^\circ$ and atomic corrugation $r=0.7$, is located at $E_{\text{VHS}}\simeq119\,\text{meV}$.
%0.246$.
The VHS is characterized by a change in the topology of the Fermi surface, a Lifschitz transition displayed in Fig. \ref{Fermi_surface_4}, where it evolves from encircling the $\bK$ and $\bK^\prime$ points to circulating around the $\bm{\Gamma}$ point.
% extended Fermi surface, as shown in Fig. \ref{Fermi_surface_4} b). 
If we increase further the Fermi energy above the saddle point, we observe a crossover from paramagnetism back to diamagnetism.
In this region the Fermi surface corresponds to a single closed line around the $\bm{\Gamma}$ point as displayed in Figs. \ref{Fermi_surface_4} c) and \ref{Fermi_surface_4} d). 
At the almost complete filling of the band, the orbital magnetic susceptibility undergoes a second sharp crossover towards paramagnetism shown in Fig. \ref{perp_OMS_angle_28} and Fig. \ref{perp_OMS_angle_alpha}, top panel. This second rise of the susceptibility is even more pronounced as the energy gap $\Delta E$ between the first two positive energy bands is small \cite{Vignale_1991}, as shown in Fig. \ref{perp_OMS_angle_alpha} by comparing the top and bottom panels. In Fig. \ref{perp_OMS_angle_alpha}, we show the evolution of the orbital susceptibility $\chi^{\perp}_{\text{OMS}}$ with the chemical potential (or doping) for different twist angles $\theta$. Already below $\theta=2^\circ$, the completely filled band is still diamagnetic despite the positive sharp increase close to the band edge.

%In order to represent the evolution of $\chi^{\perp}_{\text{OMS}}$ for different values of the twist-angle $\theta$, we plot  in the top panel of Fig. \ref{perp_OMS_angle_alpha} the orbital magnetic susceptibility as a function of the chemical potential measured in unit of the maximum of $\ep_{+\bk}$. 
%The gap $\Delta E$ between $\ep_{+\bk}$ and the second positive energy band, displayed in the bottom panel of Fig. \ref{perp_OMS_angle_alpha}, increases the value of $\chi^{\perp}_{\text{OMS}}$ at filling 1 of the low-energy band $\ep_{+\bk}$ decreases and becomes diamagnetic for $\theta=2^\circ$.

Interestingly, for a finite and small out-of-plane magnetic field, the phenomenon of magnetic breakdown occurs at the VHS. The semiclassical magnetic orbits are ill-defined at the VHS where their velocity is vanishing. At finite magnetic field, electron- and hole-like Landau levels meet at the VHS \cite{lu2014} in close relationship with the singularity we find in the orbital magnetic susceptibility.

We stress that the paramagnetic orbital susceptibility both in the region of the saddle point and in the region of the narrow gap is quite large. 
Given the typical values of the orbital magnetic susceptibility  a magnetic field of $1\text{T}$ gives rise to a magnetization density of the order of $0.02\,\mu_B/\text{nm}^2$ that for $\theta=2.3^\circ$ corresponds to $\sim 1\mu_B$ per Moir\'e unit cell. Such a large orbital magnetization may be experimentally measured by Kerr rotation microscopy \cite{Kato1910,Sih_2005}. 

In order to elucidate the origin of the large paramagnetic response at finite Fermi energy in the next Section we detail the contribution of the VHS to the orbital magnetic susceptibility $\chi^{\perp}_{\text{OMS}}$.

\subsection{The logarithmic singularity at the VHS}

\begin{figure}
\begin{center}
\includegraphics[width=0.45\textwidth]{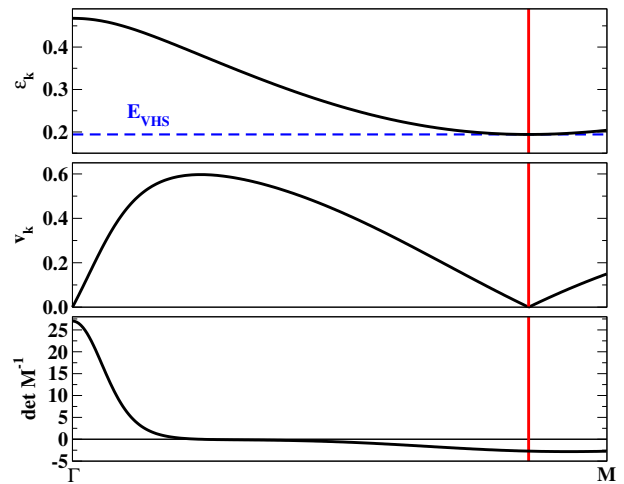}
\caption{Top: dispersion relation of the lowest positive energy band along the $\bm{\Gamma}\mathbf{M}$ line. The blue dashed line shows the energy of the VHS. Center: group velocity $\nabla_{\bk}\ep_{+\bk}$ along the $\bm{\Gamma}\mathbf{M}$ line. Bottom: determinant of the inverse mass tensor along the $\bm{\Gamma}\mathbf{M}$ line. 
The vertical red line denotes the position of the VHS, where the group velocity vanishes $|\nabla_{\bk}\ep_{+\bk}|=0$ and $\text{det}M^{-1}_{+\bk}<0$. }
\label{Saddle_point_23}
\end{center}
\end{figure}

\begin{figure}
\begin{center}
\includegraphics[width=0.45\textwidth]{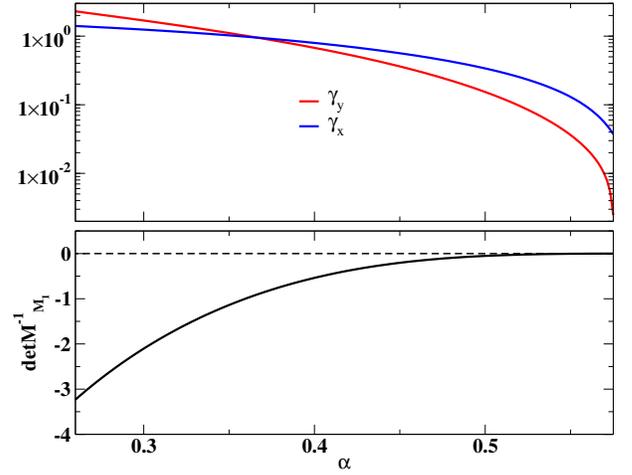}
\caption{Top: evolution of $\gamma_x$ and $\gamma_y$ as a function of the $\alpha$ in logarithmic scale. Bottom: evolution of the determinant of the inverse mass tensor as a function of $\alpha$ evaluated at the saddle point $\mathbf{M}_1$ located at the $\bm{\Gamma} \mathbf{M}$ line. The calculation has been performed by setting $r=0.7$. It is important to observe that the $\text{det}M^{-1}_\bk$ vanishes at the higher order VHS that for this value of the atomic corrugation is located at $\theta_c\simeq 1.11^\circ$. Approaching the higher-order VHS we find that $\gamma_y\to0$ while $\gamma_x$ is finite.}
\label{Saddle_point_evolution}
\end{center}
\end{figure}

The Moir\'e lattice potential gives rise to saddle points in the lowest energy bands of TBG that, above a certain critical twist-angle $\theta_c$ ($\theta>\theta_c$), are located along the $\bm{\Gamma}\mathbf{M}$ lines of the MBZ, see Fig. \ref{lattice_TBG}. 
The typical features of the band structure along the $\bm{\Gamma}\mathbf{M}$ line are shown in Fig. \ref{Saddle_point_23} where we plot the dispersion of the positive lowest energy band (top panel), the gradient (center) and the determinant of the inverse mass tensor (bottom) for $\theta=2.3^\circ$ and $r=0.7$. The vertical red line in Fig. \ref{Saddle_point_23} denotes the position of the saddle point.   
At the critical value $\theta_c\simeq 1.11^\circ$, the inverse mass tensor determinant $\text{det}M^{-1}_{+\bk}$ vanishes, see Fig. \ref{Saddle_point_evolution}. 
For $\theta<\theta_c$, two $C_{2x}$ symmetric saddle points emerge away from the $\bm{\Gamma}\mathbf{M}$ line \cite{Fu_2019}.
In proximity to the saddle point for $\theta>\theta_c$ the energy dispersion reads: 
\be
\label{En_saddle_point}
\ep_{+\mathbf{M}_1+\delta\bk}\simeq\,E_{\text{VHS}}+\left(\frac{\gamma_x}{2}k^2_x-\frac{\gamma_y}{2}k^2_y\right),
\ee   
where $\mathbf{M}_1=\mathbf{M}=-(a,0)$ and $\gamma_x,\gamma_y>0$. 
The corresponding inverse mass tensor is diagonal
\be
M^{-1}_{n\bk}=\left(\begin{array}{cc}\gamma_x & 0 \\ 0 & -\gamma_y \end{array}\right),
\ee  
and has $\text{det}M^{-1}_{+\bk} = -\gamma_x \gamma_y<0$. As it is shown in Fig. \ref{Saddle_point_evolution} the coefficients $\gamma_x$ and $\gamma_y$ depend on the twist angle $\theta$ and $\gamma_y$ vanishes at the critical angle $\theta_c\simeq 1.11^\circ$ where the van Hove singularity becomes a higher-order one.
At $\theta_c$, the TBG develops a power-law divergence in the density of states \cite{Fu_2019} underlying electronic instabilities \cite{Fu_2019_supermetal,Classen_2020,Nandkishore_2020}.

The expansion of the energy dispersion $\ep_{+\bk}$ around the saddle points $\mathbf{M}_{j+1}=\mathcal{C}^j_{3z}\mathbf{M}_1$ $(j=1,2,3)$ can be obtained from Eq. \eqn{En_saddle_point} by transforming $\bk\to\mathcal{C}^j_{3z}\bk$. Since the determinant is invariant under similarity transformation we find the singular contribution to $\chi^{\perp}_{\text{OMS}}$,
\be
\chi^{\perp,\text{VHS}}_{\text{OMS}}=\frac{e^2\,\gamma_x\gamma_y}{4\pi^2\hbar^2}\int_{<k_M} d^2\bk\,\delta\left(\ep-\frac{\gamma_x k^2_x}{2}+\frac{\gamma_y k^2_y}{2}\right),
\ee 
where $\ep$ is defined as $\ep=\mu-E_{\text{VHS}}$, and the integral over the two-dimensional $\bk$-space extends to a disk centered at $\mathbf{M}_j$ with radius $k_M$.
In the previous expression we have neglected the $\lambda^{(1)}_{n}(\bk)$ contribution \eqn{lambda_1} since the saddle point singularity is canceled by the vanishing $\partial_{k_y}\ep_{+\bk}$ factor. 
By performing straightforward calculations we find: 
\bal
&\chi^{\perp,\text{VHS}}_{\text{OMS}}=\frac{\sqrt{\gamma_x\gamma_y}\,e^2}{2\pi^2\hbar^2}\\
&\Bigg[\theta(-\ep)\theta\left(\ep+\frac{\gamma_y\,k_M^2}{2}\right)\log\frac{\sqrt{1-\frac{2\ep}{k_M^2\gamma_x}}+\sqrt{1+\frac{2\ep}{k_M^2\gamma_y}}}{\sqrt{1-\frac{2\ep}{k_M^2\gamma_x}}-\sqrt{1+\frac{2\ep}{k_M^2\gamma_y}}}\\
&+\theta(\ep)\theta\left(\frac{\gamma_x\,k_M^2}{2}-\ep\right)\log\frac{\sqrt{1-\frac{2\ep}{k_M^2\gamma_x}}+\sqrt{1+\frac{2\ep}{k_M^2\gamma_y}}}{\sqrt{1+\frac{2\ep}{k_M^2\gamma_y}}-\sqrt{1-\frac{2\ep}{k_M^2\gamma_x}}}\Bigg],
\eal
which diverges logarithmically in the $\mu\to E^{\pm}_{\text{VHS}}$ limit: 
\be
\label{OMS_perp_1_log}
\chi^{\perp,\text{VHS}}_{\text{OMS}}\simeq \frac{\sqrt{\gamma_x\gamma_y}\,e^2}{2\pi^2\hbar^2}\log\left(\frac{k^2_M}{|\ep|}\frac{\gamma_x\gamma_y}{\gamma_x+\gamma_y}\right).
\ee
Eq. \eqref{OMS_perp_1_log} confirms the divergence of the magnetic susceptibility numerically observed in Fig. \ref{perp_OMS_angle_alpha} at the VHS. It also predicts that the divergence is suppressed at the critical angle $\theta_c$. Indeed, as can be seen  in the top panel of Fig. \ref{Saddle_point_evolution}, the matrix element $\gamma_y \to 0$ despite the power-law singularity in the density of state. The origin of this suppression is the flattening of the band spectrum at the higher-order VHS corresponding to low electron velocities and thus to a weaker tendancy towards orbital ferromagnetism as already emphasized in the introduction. In other words, even though many more paramagnetic orbits are available when approaching the higher-order VHS, they rotate with a velocity suppressed by the vanishing inverse mass tensor $\gamma_x \gamma_y$.

%The logarithmic divergences is suppressed in the limit $\alpha\beta\to0$.  
%This situation is particularly relevant in TBG when the twist-angle $\theta$ approaches $\theta_c$. 

\section{The orbital magnetic instability in bilayer materials}
\label{Condon_domains_TBG}

\subsection{General criterion}

The nature of the the magnetic instability can be readily understood by promoting the classical vector potential $\mathbf{A}$ to a quantum operator.
In order to enhance the vacuum fluctuations of the electromagnetic field we embedd TBG in an optical cavity with planar geometry $L_x,L_y\gg L_z$ and $L_z\simeq 10\mu\text{m}$ where we impose periodic boundary conditions along $x$ and $y$ while metallic ones along $z$, i.e. $E_x=E_y=B_z=0$ at $z=0,\,L_z$. In the Coulomb gauge, the vector potential fulfilling the cavity boundary conditions can be expressed as follows:
\bal
\hat{\mathbf{A}}\left(\bq,z=\frac{L_z}{2}\right)&=\sum_{\bk}\sum_{n}\sum_{\sigma=1}^{2}\,\frac{\sqrt{2}e^{i\bk\cdot\br}}{\sqrt{S}}\,\alpha_{\bk n\sigma}\,\bm{\xi}_{\bk\sigma}X_{\bk n\sigma}\\
&-\mathbf{z}\sum_{\bk}\sum_{n}\frac{\sqrt{2}e^{i\bk\cdot\br}}{\sqrt{S}}\,\alpha^z_{\bk n1}X_{\bk n 1},
\eal
where $z=L_z/2$ is the vertical location of the sample, $S=L_x\,L_y$, $X_{\bk n \sigma}=(a^\dagga_{\bk n\sigma}+a^\dagger_{-\bk n\sigma})/\sqrt{2}$ and $P_{\bk n\sigma}=-i(a^\dagga_{\bk n\sigma}-a^\dagger_{-\bk n\sigma})/\sqrt{2}$ are conjugate variables $[X_{\bk n\sigma},P_{-\bk n^\prime \sigma^\prime}]=i \delta_{n,n^\prime}\delta_{\sigma,\sigma^\prime}$, and $a^\dagga_{\bk n\sigma}$ and $a^\dagger_{\bk n\sigma}$ are bosonic operators for cavity modes with wave vector $\bk=(2\pi n_x/L_x,2\pi n_y/L_y)$ $n_x,n_y\in\mathbb{Z}$, $n=0,1,\cdots$ and polarization $\sigma=1,2$.
In the previous expression we have also included the in-plane polarization vectors $\bm{\xi}_{\bk 1}=i\bk/|\bk|$ and $\bm{\xi}_{\bk 2}=i(\bm{z}\times\bk)/|\bk|$, and the coefficients: 
\bal
& \alpha_{\bk n1}=\sqrt{\frac{\hbar}{\epsilon_0\,L_z\,\omega_{\bk n}}}\,\frac{\pi n/L_z}{\sqrt{\left|\bk\right|^2+\left(\pi n/L_z\right)^2}}\sin\frac{\pi n}{L_z}, \\
& \alpha_{\bk n2 }=\sqrt{\frac{\hbar}{\epsilon\,L_z\,\omega_{\bk n}}}\sin\frac{\pi n}{L_z},\\
& \alpha^z_{\bk n1}=\sqrt{\frac{\hbar}{\epsilon\,L_z\,\omega_{\bk n}}}\,\frac{|\bk|}{\sqrt{\left|\bk\right|^2+\left(\pi n/L_z\right)^2}}\cos\frac{\pi n}{L_z},
\eal
where $\epsilon=\ep_0\,\ep_r$ with $\ep_0$ is the vacuum permittivity and $\ep_r$ relative dielectric constant, $\omega_{\bk n}=c\sqrt{|\bk|^2+(\pi n/L_z)^2}/\sqrt{\ep_r}$.
The light-matter interaction can be readily obtained by the minimal coupling scheme $\hat{\bk}\to\hat{\bk}+e\hat{\mathbf{A}}/\hbar$ and $T(\br)\to T(\br) \,e^{i ed_\perp A_z(\br)/\hbar}$ and reads: 
 \bal
 H_{\text{lm}}&=e\sqrt{\frac{2}{S}}\sum_{\bq}\sum_{n}\sum_{\sigma}\,\alpha_{\bq n\sigma}X_{ \bq n\sigma}\,\bm{\xi}_{\bq \sigma}\cdot\hat{\mathbf{J}}(-\bq)\\
 &+e\sqrt{\frac{2}{S}}\sum_{\bq}\sum_{n}\alpha^z_{\bq n1}\,X_{\bq n 1}\,\hat{J}^z(-\bq)\\
 -\frac{e^2}{S}&\sum_{\bq_1,\bq_2}\sum_{n_1,n_2}\alpha^z_{\bq_1 1n_1}X_{\bq_1n_1 1}\hat{\mathcal{T}}^{z,z}(-\bq_1-\bq_2)\\
 &\alpha^z_{\bq_2 1n_2}X_{\bq_2 n_2 1},
 \eal
 where the first term describes the coupling between the in-plane components of the $\sigma=1,2$ modes, while the latter the coupling to the component of the mode $\sigma=1$ along the $\bm{z}$ axis. Finally, we have the energy of the cavity modes that reads $H_{\text{cav}}=\sum_{\bq,n,\sigma}\hbar\omega_{\bk n}(a^\dagger_{\bk n\sigma}a^\dagga_{\bk n\sigma}+1/2)$.

We derive the magnetic instability conditions by the vanishing of the lowest polariton frequency as detailed in the Appendix~\ref{cavity_MODES}. 
We find that cavity photons condense to form an out-of-plane spatially modulated magnetic field when
\be
\label{CD_out_plane}
\frac{\chi^\perp_{\text{OMS}}(\bq)}{\chi_0(L_z)}\ge\frac{1}{qL_z\,\tanh(q L_z/2)/2}.
\ee
On the other hand, the condition for the occurrence of photon condensation to an in-plane magnetic field reads:
\be
\label{CD_in_plane}
\frac{\chi^\parallel_{\text{OMS}}(\bq)}{\chi_0(L_z)}\ge\frac{1}{1+\frac{qL_z}{4\pi}\coth\frac{qL_z}{2}+\frac{(q\,L_z)^2}{8\sinh(qL_z/2)^2}},
\ee
where the quantum of the orbital magnetic susceptibility is $\chi_0(L_z)=L_z\,e^2 c/(4\pi \alpha_{EM}\hbar)$ and depends linearly on the length $L_z$ of the cavity. 
Eq.~\eqref{CD_out_plane} recovers exactly the criterion derived in Ref.~\cite{Andolina_2020,Basko_PRL2019}, where the instability was identified with the formation of Condon domains \cite{Condon_1966,Azbel_1968,Holstein_1973,Markiewicz_1985,Quinn1985,Wyder_1998} of spontaneous magnetization. This transition is also closely connected to superradiant quantum phase transitions \cite{Basko_PRL2019,Andolina_2020,Guerci_2020} in optical cavities. 
Interestingly, Eq.~\eqref{CD_in_plane} is the condition for the occurrence of photon condensation for a magnetic field lying in the plane of the top and bottom layers. 
This situation can be realized in bilayer materials where ground state interlayer currents loop induce an in-plane magnetic field.
Motivated by the experimental observation of ferromagnetic splitting at half-filling of the VHS in TBG \cite{Liu_2019,Kerelsky_2019,Choi_2019} perpendicularly to the bilayer plane, we discuss in the next Section the possibility of an orbital magnetic instability in TBG when embedded in an optical cavity and the resulting symmetry-broken phase.

\subsection{Out-of-plane orbital ferromagnet}

We now investigate whether the criterion in Eq.~\eqn{CD_out_plane} can be satisfied in TBG. 
To this aim we consider the quasihomogenous limit~\cite{Andolina_2020} where the wavevector $|\bq|$ is much smaller than the electronic length scale given by the inverse of the size of the Moir\'e unit cell $k_\theta\sim10^9-10^8\,\text{m}^{-1}$ but still larger or equal than the vertical size of the cavity $L_z\simeq 10\,\mu\text{m}$, $1/L_z\lesssim |\bq|\ll k_\theta$. In this regime we can safely replace $\chi^{\perp}_{\text{OMS}}(\bq)$ with $\chi^{\perp}_{\text{OMS}}=\chi^{\perp}_{\text{OMS}}(\bq\to0)$ and the right hand side of Eq.~\eqref{CD_out_plane} with $\sim1$, so that we have: 
\be
\label{CD_out_plane_1}
\chi^{\perp}_{\text{OMS}}/\chi_0(L_z)\gtrsim1.
\ee  

Even though the divergence of Eq.~\eqref{OMS_perp_1_log} suggests that the instability~\eqref{CD_out_plane_1} takes place in TBG, the effect is very weak in practice. For a perpendicular width $L_z\simeq 10\mu\text{m}$, the value of the quantum of the orbital susceptibility is $\chi_{0}(L_z)\simeq\,8.2\,10^5\mu_B/\text{T nm}^2$. 
This is several orders of magnitude larger than the typical values observed in Figs. \ref{perp_OMS_angle_28} and \ref{perp_OMS_angle_alpha}. Close to the divergence at the VHS, the condition for the instability, obtained from Eqs.~\eqref{OMS_perp_1_log} and~\eqref{CD_out_plane_1}, is  
\be
\label{energy_range}
|\ep|\le \frac{k^2_M \gamma_x\gamma_y}{\gamma_x+\gamma_y}\exp\left(-\frac{\pi\hbar c L_z}{2\alpha_{EM}\,\sqrt{\gamma_x\gamma_y}}\right). 
\ee
with $\ep=\mu-E_{\text{VHS}}$. Since $\sqrt{\gamma_x\gamma_y}$ has the dimension of an energy per length square we have $\pi\hbar c L_z/(2\alpha_{EM}\sqrt{\gamma_x\gamma_y})\sim (L_z k_\theta) c/(v_0\alpha_{EM}\sqrt{\gamma_x\gamma_y})$, where $v_0$ is the typical Fermi velocity of graphene. 
In the interval of twist angles $\theta=2^\circ-3^\circ$, $L_z k_\theta\sim 10^3$, $\sqrt{\gamma_x\gamma_y}\sim 1$ and we obtain in Eq.~\eqref{energy_range} an exponential suppression with the factor $L_z k_\theta\,c/(\alpha_{EM}\,v_0)\sim 10^3\,(c/v_0)^2\sim 10^7$.

Condon phases, and the criterion~\eqref{CD_out_plane} in the quasihomogenous limit, correspond to an instability where the magnetic induction becomes a multivalued function of the magnetic field, leading to two coexisting phases with different orbital magnetizations \cite{Condon_1966,Azbel_1968,Holstein_1973,Markiewicz_1985,Quinn1985,Wyder_1998}. 
So far, Condon domains have been measured experimentally under moderate magnetic fields. Here we consider an alternative scenario where they would appear in the absence of external magnetic field, only triggered by a large paramagnetic susceptibility. 
We examine the ground state properties with a mean-field decoupling of the unscreened current-current interaction obtained by integrating out the whole spectrum of $\sigma=2$ cavity modes: 
\be
H_{JJ}=-\frac{\pi\alpha_{EM}\hbar L_z}{2cS}\sum_{\bq}g_{q}\,\hat{J}_{T}(-\bq)\hat{J}_{T}(\bq),
\ee
where $g_q=2\tanh(|\bq|\,L_z/2)/L_z|\bq|$ and $\hat{J}_{T}(\bq)=\bm{\xi}_{-\bq 2}\cdot\hat{\mathbf{J}}(\bq)$ with $\hat{\mathbf{J}}(\bq)$ defined in Eq. \eqn{charge_dipolar}.
A non-vanishing transverse current operator $\langle \hat{J}_{T}(\bq)\rangle$ characterizes the ground state. It spontaneously breaks time reversal symmetry as well as the $C_{2z}T$ symmetry. The average current operator corresponds to an orbital magnetization through
\begin{equation}
\langle \hat{J}_{T}(\bq)\rangle=\frac{q \, S}{e} M^{\text{orb}}_{z}(\bq),
\end{equation}
$S$ is the surface of the TBG sample. 
To express the Hartree-Fock Hamiltonian in the homogeneous limit $1/L_z\lesssim |\bq|\ll k_\theta$, we find it convenient to introduce the gauge field $\bm{a}(\br)$ such that
\be
M^{\text{orb}}_z(\br)=\epsilon_{ijz} \partial_i a_j(\br),
\ee
or in Fourier space $M^{\text{orb}}_z(\bq)=i\epsilon_{lsz}\,q_l  a_s(\bq)$. We reach the quasihomogeneous limit with the choice $\bq=q\,\bm{x}$ and $q\sim1/L_z$. With this gauge choice, we find the Hartree-Fock form for the current-current interaction
\begin{equation}
\begin{split}
H^{HF}_{JJ}=&\frac{2g\pi\alpha_{EM}\hbar}{c\,L_z\,e^2}\int d\br\, \Big[ 2 e \hat{J}_y(x)\,a_y(x) \\
&+ ( M^{\text{orb}}_z(\br) )^2 \Big],
\end{split}
\end{equation}
where $g=2\tanh(1)$.
It is transparent with this writting that the orbital magnetization plays the role of an effective magnetic field applied to the TBG and the gauge field $\bm{a}(\br)$ is its vector potential. For an orbital magnetization varying on the length scale $\sim L_z$ much larger that the electronic one $\sim 1/k_\theta$, the gauge field can be chosen as $\bm{a}(\br)=x\, M^{\text{orb}}_z\bm{y}$ and the minimization of the Hartree-Fock energy yields the familiar expression
\be
M^{\text{orb}}_z=-\frac{e}{S} \langle \hat{x} \,\hat{J}_y \rangle,
\ee 
identifying the orbital magnetization as a sum over angular orbital momenta~\cite{Resta_2005}. In summary, the Hartree-Fock approach introduces an effective magnetic field seen by the electrons of TBG. The resulting ground state exhibits Landau levels similarly to the Hofstadter spectrum~\cite{Lu2020_multiple,Herzog_2020,Lian_2020_LLs}, with an orbital magnetization self-consistently determining the effective magnetic field.

\section{The OMS to an in-plane magnetic field in TBG}
\label{OMS_in_plane}

\subsection{General expression}

In terms of the dipolar current operator $\hat{\bm{\xi}}$ defined in Eq. \eqn{charge_dipolar} the orbital magnetic response to an in-plane magnetic field \eqn{OMS_paral} with finite wavevector modulation $\bq$ reads: 
\bal
\label{OMS_paral_1}
\chi^{\parallel}_{\text{OMS}}(\bq)&=-g_sg_v e^2\sum_{n,n^\prime}\int_{\text{MBZ}}\frac{d^2\bk}{4\pi^2}\\
&\frac{|\bra{u_{n\bk}}\xi^y\ket{u_{n^\prime\bk+\bq}}|^2}{\ep_{n\bk}-\ep_{n^\prime\bk+\bq}+i0^+}[f(\ep_{n\bk})-f(\ep_{n^\prime\bk+\bq})],
\eal
where $\xi^y_{\bQ\alpha,\bQ^\prime\beta}=v_0d_\perp\delta_{\bQ,\bQ^\prime}\gamma(\bQ)\,\sigma^y_{\alpha,\beta}/2$ with $\gamma(\bQ)=+1$ if $\bQ\in\mathcal{Q}_2$ while $\gamma(\bQ)=-1$ if $\bQ\in\mathcal{Q}_1$.
Similarly to Sec. \ref{OMS_out_of_plane} we split $\chi^{\parallel}_{\text{OMS}}(\bq)$ in the sum of the intraband ($n=n^\prime$) and the interband ($n\neq n^\prime$) contribution: $\chi^{\parallel}_{\text{OMS}}(\bq)=\chi^{\parallel,\text{intra}}_{\text{OMS}}(\bq)+\chi^{\parallel,\text{inter}}_{\text{OMS}}(\bq)$.
By taking the $\bq\to0$ limit we find that the intraband contribution reduces to the semiclassical result:
\be
\label{OMS_paral_intra}
\chi^{\parallel,\text{intra}}_{\text{OMS}}=\frac{g_sg_v e^2}{4\pi^2}\sum_n\int_{\mathcal{S}_{n,\mu}}dl\frac{|\bra{u_{n\bk}}\xi^y\ket{u_{n\bk}}|^2}{|\nabla_\bk\ep_{n\bk}|},
\ee
where $\mathcal{S}_{n,\mu}$  constant energy contour $\ep_{n\bk}=\mu$ in the $n$th band. 
Interestingly, this result can be recovered from a purely semiclassical computation following the reasoning in Ref.\cite{DTSon_2020}. We use a Boltzmann equation approach to describe the motion of electrons. We remark that the use of the Boltzmann equation relies of the mean free path $l=v\,\tau$ (scattering time times velocity) being much larger than the electronic length scale $l_{\text{matter}}$, $l\gg l_{\text{matter}}$. In the specific case of TBG, we require that $l$ is much larger than the linear size of the Moir\'e unit cell $L_\theta=4\pi/3k_\theta$ and also than the distance $L_\text{jump}=L_\theta/\alpha=2\hbar v_0/3w$ over which one electron tunnels from one layer to the other. In the magic angle regime $L_\text{jump}\sim L_\theta \simeq O(10\,\text{nm})$, so requiring that $l\gg L_{\text{jump}}$ is essentially the same condition as for using the band structure of TBG.
Under this assumption we can introduce a non equilibrium distribution $f^{\text{neq}}_{n\bk}$ and the Boltzmann equation in an uniform in-plane magnetic field reads:
\begin{equation}\label{boltzmann}
\partial_t\,f^{\text{neq}}_{n\bk}=-(f^{\text{neq}}_{n\bk}-f(E_{n\bk}))/\tau
\end{equation}
with the distribution function $f^{\text{neq}}_{n\bk}$, $n$ the index running over the spectrum of TBG, the impurity scattering rate $\tau$ and $f(\ep)$ is the Fermi-Dirac distribution. 
In Eq. \eqn{boltzmann} the quasiparticle energy takes the form $E_{n\bk}=\ep_{n\bk}-\bm{M}^{\text{orb}}_{n,\parallel}(\bk)\cdot\bm{B}$ \cite{Niu_1995,Niu_1996,Niu_1999} where the in-plane orbital magnetization is $\bm{M}^{\text{orb}}_{n,\parallel}(\bk)=\bra{u_{n\bk}}e\,\pmb{\xi}\times\mathbf{z}\ket{u_{n\bk}}$. 
We linearize the solution to Eq. \eqref{boltzmann} as $f^{\text{neq}}_{n\bk}=f(\ep_{n\bk})+\delta f_{n\bk}$, finding $(1-i\omega\tau)\delta f_{n\bk}=-\bm{M}^{\text{orb}}_{n,\parallel}(\bk)\cdot\bm{B}\,[\partial_\ep f(\ep)]|_{\ep_\bk}$ which gives in the stationary limit $\omega\tau\ll1$ the orbital magnetization (from now on we suppress the band index $n$):
%Therefore, a Boltzmann equation approach that accounts only the low-energy quasiparticle motion \cite{Niu_1995,Niu_1996,Niu_1999,DTSon_2020}, i.e. the contribution in Eq. \eqn{OMS_paral_intra}, would necessarily miss this effect.
%Indeed, in this description the energy of the quasiparticles in an in-plane magnetic field has the form $E_\bk=\ep_\bk-\bm{M}^{\text{orb}}_{\parallel}\cdot\bm{B}$. Within the relaxation-time approximation we find that the Boltzmann equation for the distribution function $f^{\text{neq}}_\bk$ reads $\partial_t\,f^{\text{neq}}_\bk=-(f^{\text{neq}}_\bk-f(E_\bk))/\tau$ where $f(\ep)$ is the Fermi-Dirac distribution. By linearizing the equation, $f^{\text{neq}}_\bk=f(E_\bk)+\delta f_\bq$, we find $\delta f_\bk=-\bm{M}^{\text{orb}}_{\parallel}\cdot\bm{B}\,\partial_\ep f(\ep_\bk)$, so that orbital magnetization $\bm{M}^{\text{orb}}_{\parallel}$ induced by the in-plane magnetic field is given by: 
\bal
\label{M_orb_linearized}
M^{\text{orb}}_i&=g_v\,g_s\,\int\frac{d^2\bk}{(2\pi)^2}\,e\ep_{ijz}\xi_j\,\delta f_{\bk}\\
&=g_v\,g_s\,e^2\langle \xi_j\,\xi_j \ep_{ijz}\ep_{ijz}\rangle B_i,
\eal
where $i=x,y$, $\ep_{ijk}$ is the Levi-Civita tensor, and $\langle \cdot \rangle$ is the average over the constant energy line $\mathcal{S}_\mu$, $\langle A \rangle=(1/4\pi^2)\int_{\mathcal{S}_\mu} dl \,A/|\nabla \ep_\bk|$. By summing over the spectrum of TBG Eq. \eqn{M_orb_linearized} reads: 
\be
M^{\text{orb}}_i=\frac{g_sg_v e^2}{4\pi^2}\sum_n\int_{\mathcal{S}_{n,\mu}}dl\frac{|\bra{u_{n\bk}}\xi^j\,\ep_{ijz}\ket{u_{n\bk}}|^2}{|\nabla_\bk\ep_{n\bk}|}\,B_i,
\ee 
which coincides with the intraband contribution in Eq.~\eqref{OMS_paral_intra}.
% We observe that the linear response coefficient in Eq. \eqn{M_orb_linearized} coincides with the intraband contribution in Eq. \eqn{OMS_paral_intra} and neglects the interband one \eqn{OMS_paral_inter}.

In contrast with the intraband contribution evaluated on the Fermi surface, the interband term involves a summation over the whole spectrum. It requires an ultraviolet regularization, already formulated in Refs. \cite{Castro_Neto_fsumrule_2008,Stauber_2013}, which can alternatively be eluded with a different choice of gauge, see appendices~\ref{in_plane_Aperp_gauge} and~\ref{reg_interlayer}. The regularization is implemented with an energy cutoff $\Lambda$  in the particle-hole excitation spectrum, and
\bal
\label{OMS_paral_inter_1}
\overline{\chi}^{\parallel,\text{inter}}_{\text{OMS}}(\Lambda)=&-\frac{g_sg_ve^2}{4\pi^2}\sum_{n,n^\prime}^{n\neq n^\prime}\int_{\text{MBZ}}d^2\bk\,\theta[\Lambda-|\ep_{n\bk}-\ep_{n^\prime\bk}|] \\
&\frac{ |\bra{u_{n\bk}}\xi^y\ket{u_{n^\prime\bk}}|^2}{\ep_{n\bk}-\ep_{n^\prime\bk}}[f(\ep_{n\bk})-f(\ep_{n^\prime\bk})],
\eal
where $\theta(\ep)$ is the Heaviside step function. The interband contribution is then obtained by subtracting the same quantity for decoupled layers, 
\be
\label{OMS_paral_inter_2}
\chi^{\parallel,\text{inter}}_{\text{OMS}} =\overline{\chi}^{\parallel,\text{inter}}_{\text{OMS}}(\Lambda)-\overline{\chi}^{\parallel,\text{inter}}_{\text{OMS}}(\Lambda,\alpha=0),
\ee
and sending the cutoff energy  $\Lambda$ to infinity. In practice, convergence is already achieved for $\Lambda/\hbar v_0k_\theta\sim4-5$ as shown in Fig. \ref{in_plane_UV_cutoff} of Appendix \ref{reg_interlayer}. As also shown in Appendix~\ref{reg_interlayer}, the result~\eqref{OMS_paral_inter_2} coincides numerically with the expression~\eqn{OMS} obtained in the alternative gauge where no regularization is needed.

We finally stress that, since the intraband contribution~\eqn{OMS_paral_intra} is always positive, i.e. paramagnetic, the diamagnetic or paramagnetic character of $\chi^{\parallel}_{\text{OMS}}$ depends on the sign and magnitude of the interband contribution $\chi^{\parallel,\text{inter}}_{\text{OMS}}$.

\begin{figure}
\begin{center}
\includegraphics[width=0.48\textwidth]{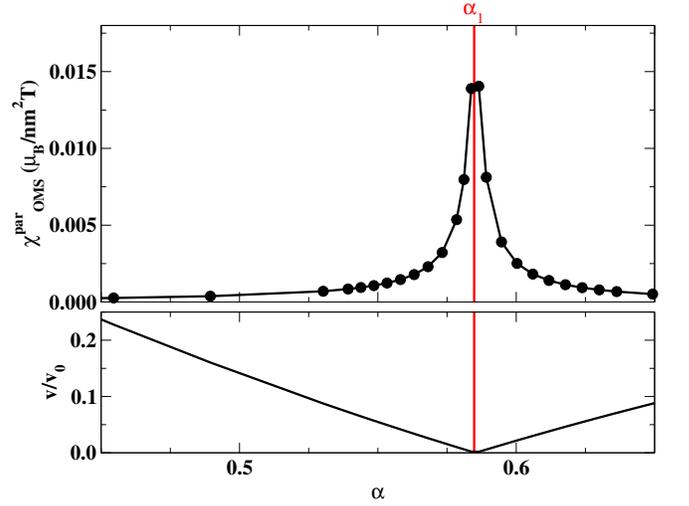}
\caption{Top: black data show the orbital magnetic susceptibility as a function of the interlayer coupling $\alpha$ for $\mu=0$ and $r=0.2$. 
Bottom: renormalization of the low-energy electron's velocity as a function of $\alpha$ for $r=0.2$. 
The vertical red line denotes the first magic angle $\alpha_1$ located at $\theta_1\simeq1.09^\circ$.
The value of $\chi^\parallel_{\text{OMS}}$ is measure in unit of $\mu_B/\text{nm}^2\,\text{T}$, where $\mu_B$ is the Bohr magneton, $\text{T}$ is Tesla and refers to the intensity of the magnetic field.}
\label{par_OMS_alpha}
\end{center}
\end{figure}

\subsection{Results at charge neutrality}

We focus our discussion on charge neutrality where the chemical potential $\mu=0$ vanishes and the Fermi surface reduces to a set of isolated points at $\bK$ and $\bK^\prime$. As a result, the intraband term \eqref{OMS_paral_intra} vanishes and we are left with the interband contribution only.
For the same reason the in-plane Pauli susceptibility $\chi_{\text{spin}}=g_s\,g_v\mu_B\,\rho(\mu=0)=0$ is vanishing and the only response at $\mu=0$ to $\mathbf{B}_\parallel$ is given by the orbital motion of the electrons.
%At charge neutrality, {\it i.e.} $\mu=0$ for $H_0$ in Eq. \eqn{BM_TBG}, the evolution of the orbital magnetic susceptibility to an in-plane magnetic field as a function of the twist-angle $\theta$ is shown in the top panel of Fig. \ref{par_OMS_alpha}.
The evolution of the orbital magnetic susceptibility to an in-plane magnetic field as a function of the twist-angle $\theta$ is shown in the top panel of Fig. \ref{par_OMS_alpha}.
%For $\mu=0$, the Fermi line shrinks to a set of isolated points at $\bK$ and $\bK^\prime$.
%Therefore, the intraband contribution \eqn{OMS_paral_intra} vanishes and the value of $\chi^{\parallel}_{\text{OMS}}$ is determined by the interband contribution \eqn{OMS_paral_inter_2}. 
Interestingly, we find a paramagnetic response which is strongly enhanced as one approaches the first magic angle, developing a logarithmic singularity in $\chi^{\parallel}_{\text{OMS}}$. Even if the light-matter coupling to the in-plane magnetic field is multiplied by small the dimensionless parameter $\left(d_\perp\,k_\theta\right)^2$ the magnetic response is not negligible. In particular, in the region close to the first magic angle we find for a magnetic field of $1\text{T}$ an orbital magnetization of the order of $0.01\,\mu_{B}/\text{nm}^2$, which is the same order of magnitude observed in the out-of-plane susceptibility close to the VHS. We notice that a magnetization $0.01\,\mu_{B}/\text{nm}^2$ corresponds for $\theta=1.09^\circ$ to $\sim2\,\mu_B$ per Moir\'e unit cell.

\begin{figure}
\begin{center}
\includegraphics[width=0.46\textwidth]{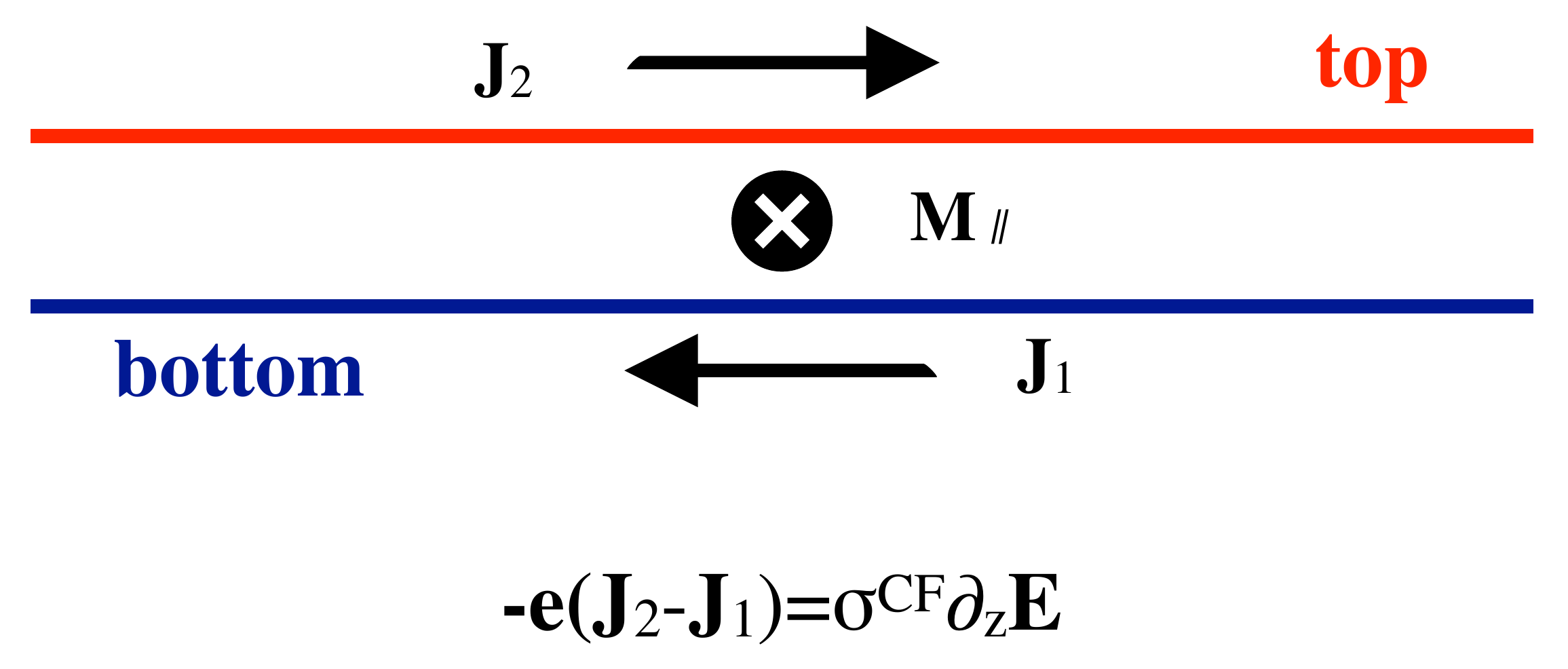}
\caption{The counterflow conductivity is defined as the transport coefficient that relates the current imbalance $\bm{J}_2-\bm{J}_1$ to the vertical gradient of the electric field $\partial_z\bm{E}$. 
A finite in-plane orbital magnetization $\bm{M}_\parallel$ implies a steady-state dipolar current $\bm{\xi}$ which underlines a close relation between $\sigma^{\text{CF}}$ and $\chi^{\parallel}_{OMS}$.}
\label{sketch_CF}
\end{center}
\end{figure}
The close relation between the in-plane orbital magnetization and the dipolar current implies a correspondence between $\chi^{\parallel}_{\text{OMS}}$ and the counterflow conductivity $\sigma^{\text{CF}}$ which is sketched in Fig. \ref{sketch_CF} and detailed in Appendix \ref{counterflow_OMS}.
In particular, as already observed in Refs. \cite{Stauber_2018,Stauber_2018_new} the two quantities are proportional  at charge neutrality and the paramagnetic orbital response implies a negative counterflow conductivity in the absence of low-energy carriers.
This intriguing effect eludes a semiclassical description since the corresponding semiclassical contribution is vanishing at $\mu=0$. It depends on the contrary on all filled electronic states.

In order to get analytical insight on the logarithmic singularity at the first magic angle we provide in the next section a detailed analysis of the contribution to $\chi^{\parallel}_{\text{OMS}}$ from the $\bK$ and $\bK^\prime$ points in the MBZ.

\subsubsection*{Contribution from the Dirac points to the in-plane OMS}

The origin of the logarithmic singularity at the magic angle observed in Fig. \ref{par_OMS_alpha} can be understood by looking at $\bk\cdot\bp$ Hamiltonian \cite{Hejazi_2019}: 
\bal
\label{BM_TBG_kp}
H_{\bk\cdot\bp}(\bK+\delta\bk)=&v\,\delta\bk\cdot\bm{\sigma}+\eta(\delta k^2_y-\delta k^2_x)\,\sigma^y\\
&-2\eta\,\delta k_x\,\delta k_y\,\sigma^x.
\eal
The previous Hamiltonian has been obtained by projecting the Eq. \eqn{BM_TBG} on the zero-energy doublet $\{\ket{\phi_{\omega}},\ket{\phi_{\omega^*}}\}$, which are simultaneous eigenstates of $H_0$ at $\bK$ and of $\mathcal{C}_{3z}$ with eigenvalues $\omega=\exp(2\pi i/3)$ and $\omega^*$. 
The divergent contribution to the orbital magnetic susceptibility $\chi^{\parallel}_{\text{OMS}}$ is given by:
\be
\label{OMS_paral_1_log}
\chi^{\parallel}_{\text{OMS}}\simeq\frac{2g_sg_ve^2}{\pi^2}\int_{<k_M}d^2\bk\,\frac{|\bra{\phi_{-\bk}}\xi^y\ket{\phi_{+\bk}}|^2}{\gamma(\bk)},
\ee
where the factor $2$ takes into account the contribution of the two Dirac points in the MBZ, the integral is extended on a circular region at $\bK$ with radius $k_M$, $\gamma(\bk)$ is the dispersion of the $\bk\cdot\bp$ Hamiltonian \eqn{BM_TBG_kp}:
\be
\gamma(\bk)=\sqrt{(v k_x-2\eta k_x k_y)^2+(v k_y+\eta( k^2_y- k^2_x))^2},
\ee
 the eigenstates $\ket{\phi_{\pm\bk}}$ are
\be
\ket{\phi_{\pm\bk}}=(\ket{\phi_\omega}\pm e^{i\Phi(\bk)}\ket{\phi_{\omega^*}})/\sqrt{2},
\ee 
with $\tan\Phi(\bk)=(v\sin\theta-\eta|\bk|\cos2\theta)/(v\cos\theta-\eta|\bk|\sin2\theta)$ and $\theta=\arctan k_y/k_x$. 
\begin{figure}
\begin{center}
\includegraphics[width=0.45\textwidth]{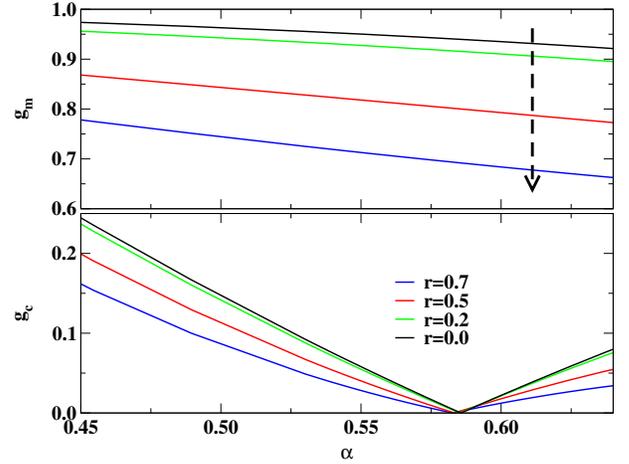}
\caption{Top: the dimensionless expectation value $g_m$ as a function of $\alpha$ for various $r$. 
Bottom:  the dimensionless expectation value $g_c$ as a function of $\alpha$ for various $r$. 
Differently from $g_c$ which vanishes at the magic angle $g_m$ is finite and gives rise to the logarithmic singularity in the $\chi^{\parallel}_{\text{OMS}}$ at the magic angle. The dashed arrow shows that $g_m$ decreases with increasing $r=w_{AA}/w_{AB}$ and takes its maximum value in the chiral limit $r=0$.}
\label{g_coupling}
\end{center}
\end{figure}
The dipole current matrix element in Eq. \eqn{OMS_paral_1_log} can be easily obtained as: 
\be
\bra{\phi_{-\bk}}\xi^y\ket{\phi_{+\bk}}=-id_\perp v_0\,g_{m}\sin\Phi(\bk)/2,
\ee
where $g_m=i\bra{\phi_\omega}(J_{y,2}-J_{y,1})\ket{\phi_{\omega^*}}/v_0$ is evaluated in the top panel of Fig. \ref{g_coupling} as a function of the twist-angle $\alpha$ for two different values of $r$. 
In the bottom panel we also show the expectation value $g_c=i\bra{\phi_\omega}(J_{y,2}+J_{y,1})\ket{\phi_{\omega^*}}/v_0$, where the operator $J_{i,l}$ is defined in Eq. \eqn{current_layer}. 
It is crucial to observe that differently from $g_c$, which vanishes at the magic angle, $g_m$ is always finite and increases by decreasing $r$.  
At the magic angle ($v=0$) we find: 
\be
\chi^{\parallel}_{\text{OMS}}\simeq\frac{2e^2}{\pi}\frac{(d_\perp v_0\,g_m)^2}{\eta}\log\frac{k_M}{\Delta},
\ee
where $\Delta$ is an infrared cutoff. 
Interestingly, the prefactor that multiplies the log is inversely proportional to the curvature $\eta$ which, close to the magic angle, is very small and amplifies the in-plane orbital magnetic susceptibility. 
The effect is even more enhanced in the chiral limit ($r=w_{AA}/w_{AB}=0$) where at the magic angle the low energy states are dispersionless \cite{Vishwanath_2018,Xi_Dai_PRB_2019,becker2020spectral,wang2020chiral}.

\subsection{Results away from charge neutrality}

\begin{figure}
\begin{center}
\includegraphics[width=0.45\textwidth]{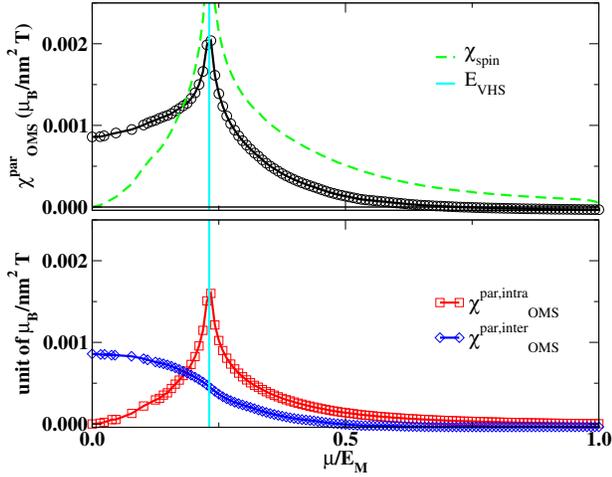}
\caption{Top: black data show the in-plane orbital magnetic susceptibility as a function of the Fermi energy $\mu/E_M$ ($E_M$ is the maximum of the positive energy flat band $E_M=12.8\,\text{meV}$) for $\theta=1.2^\circ$ $(\alpha\simeq0.53)$ and $r=0.7$. 
The black solid horizontal helps to distinguish the diamagnetic $\chi^\parallel_{\text{OMS}}<0$ from the paramagnetic $\chi^\parallel_{\text{OMS}}>0$ regime. 
The cyan vertical line denotes the energy of the VHS.
Bottom: red and blue data correspond to the intraband and interband contributions to $\chi^\parallel_{\text{OMS}}$, respectively. 
For comparison, we show in green the spin magnetic susceptibility $\chi_{\text{spin}}$.
In this case the value of $\chi_{\text{spin}}$ close to the VHS is around 2 times larger than $\chi^\parallel_{\text{OMS}}$.
The value of $\chi^\parallel_{\text{OMS}}$ is measured in unit of $\mu_B/\text{nm}^2\,\text{T}$.}
\label{par_OMS_away_CNP}
\end{center}
\end{figure}

At finite doping the Fermi surface contribution \eqn{OMS_paral_intra} does not vanish anymore and the in-plane orbital magnetic susceptibility is given by the sum of the intraband and interband contributions. 
The top panel in Fig. \ref{par_OMS_away_CNP} shows our result for the orbital magnetic susceptibility $\chi^\parallel_{\text{OMS}}$ at twist-angle $\theta=1.2^\circ$ ($\alpha\simeq0.53$) as a function of the Fermi energy $\mu/E_M$ where $E_M$ is the bandwidth of the lowest positive energy band. 
As we move away from the charge neutrality point the intraband contribution, red line in the bottom panel of Fig. \ref{par_OMS_away_CNP}, grows linearly with the chemical potential $\mu$, while the interband one, blue line in the bottom panel of Fig. \ref{par_OMS_away_CNP}, starts to decrease. 
As shown in Fig. \ref{par_OMS_away_CNP} the paramagnetic response diverges logarithmically at the energy of the VHS, which is reported as the cyan vertical line in Fig. \ref{par_OMS_away_CNP}. 
Differently to $\chi^\perp_{\text{OMS}}$, $\chi^\parallel_{\text{OMS}}$ is enhanced by approaching the flat-band regime and the logarithmic singularity becomes a power-law divergence at the higher-order VHS ($\theta\simeq1.11^\circ$ for $r=0.7$).  
The different behaviour can be easily understood by observing that the dipole current matrix element $|\bra{u_{n\bk}}\xi^y\ket{u_{n\bk}}|^2$ in Eq. \eqn{OMS_paral_intra} does not vanish at the higher-order VHS. As a result, for this specific value of the twist-angle the in-plane orbital response goes like the density of states $\chi^\parallel_{\text{OMS}}(\ep)\propto |\ep|^{-1/4}$ for $\ep=\mu-E_{\text{VHS}}\to0$.  
By further increasing the chemical potential we observe a crossover around $\mu/E_M\simeq 0.7$ from paramagnetism to weak diamagnetism. 
   
\section{Conclusions}
\label{Con}

In conclusion, we studied the orbital magnetic response of TBG in the absence of interactions.
The response to an out-of-plane magnetic field is diamagnetic at small chemical potential. 
As the chemical potential increases in the lowest positive energy band of TBG, we observe a crossover from diamagnetism to paramagnetism with a logarithmic paramagnetic singularity at the VHS. This logarithmic singularity is stronger for larger twist angles and becomes suppressed when approaching the higher-order VHS where the spectrum flattens.
Above the VHS, a crossover back to diamagnetism is obtained.
If the band gap between the first and the second positive energy bands is narrow, a large paramagnetic response is obtained at the top edge of the first band.
The divergence in the orbital magnetic response at the VHS gives rise to Condon domain phases. We described it using a mean-field picture in which an effective self-consistent magnetic field emerges with its corresponding Landau levels, finite Chern numbers and a finite orbital magnetization. The domain of existence of this Condon domain was however found to be exceedingly small, suppressed by the ratio of velocities to the speed of light.

In addition we also investigated the orbital response to an in-plane magnetic field which originates from the interlayer motion of the electrons. 
Interestingly, we found a paramagnetic response at charge neutrality which diverges logarithmically at the magic angle. 
This response originates from a large set of occupied bands. 
It is related to a negative counterflow response of the bilayer system which has no semiclassical equivalent.
Away from charge neutrality the in-plane orbital susceptibility shows a logarithmic paramagnetic singularity at the VHS. 
Differently from the out-of-plane response, the in-plane one is enhanced when approaching the first magic-angle region.

\section{Acknowledgments}
We acknowledge discussions with Marco Polini, Marcello Andolina and Francesco Pellegrino that stimulated this work. CM also acknowledges fruitful discussions with Kry\v{s}tof Kol\'{a}\v{r} and Felix von Oppen.
This work was supported by the French National Research Agency (project  SIMCIRCUIT, ANR-18-CE47-0014-01).

\appendix 
\section{a different gauge for the in-plane magnetic field}
\label{in_plane_Aperp_gauge}

In Sec. \ref{coupling_external_field} the in-plane magnetic field $\mathbf{B}_\parallel$ has been introduced by an in-plane gauge field that in the Coulomb gauge reads $\mathbf{A}(z)=z\,(B_y,-B_x,0)$. 
Interestingly, the minimal substitution can be equivalently performed by considering $\mathbf{A}(\br)=\mathbf{z}\,A_z(\br)$, where $A_z(\br)$ depends on the in-plane coordinates only.
In this gauge the orbital effect of the in-plane magnetic field can be incorporated by the Peierls substitution: 
\be
\label{different_gauge}
T(\br)\to T(\br)\,e^{ie d_\perp A_z(\br)/\hbar},
\ee
where the exponent $ed_\perp A_z(\br)/\hbar$ is the phase accumulated in an interlayer hopping process.
Because we are interested in the linear we expand Eq. \eqn{different_gauge} to second order in the $A_z(\br)$ field 
\bal
\delta H=&\frac{e\,d_\perp}{\hbar}\left(\begin{array}{cc}0 & i\,T(\br) \\ -i\,T^\dagger(\br) & 0 \end{array}\right)A_{z}(\br)\\
&-\frac{1}{2}\left(\frac{e\,d_\perp}{\hbar}\right)^2A_{z}(\br)\left(\begin{array}{cc}0 & T(\br) \\ T^\dagger(\br) & 0 \end{array}\right)A_{z}(\br).
\eal
Within the second quantization formalism we have:
\bal
&\delta H = e\sum_{\bq}\,J_{p,z}(\bq)\,A_z(-\bq)\\
&-e^2\,\sum_{\bq_1,\bq_2}\,A_z(-\bq_1)\,\mathcal{T}^{z,z}(\bq_1+\bq_2)\,A_{z}(-\bq_2)/2,
\eal
where 
\be
J_{p,z}(\bq)=\sum_{\bk\in\text{MBZ}}\sum_{\bQ\bQ^\prime\in\mathcal{Q}_{1/2}}\sum_{\alpha,\beta}\,c^\dagger_{\bk,\bQ,\alpha}\,J^{p,z}_{\bQ\alpha,\bQ^\prime\beta}\,c^\dagga_{\bk+\bq,\bQ^\prime,\beta}
\ee
with
\be
J^{p,z}_{\bQ\alpha,\bQ^\prime\beta}=\frac{i\,d_\perp}{\hbar}\sum_{j=1}^{3}T^j_{\alpha\beta}\,\left[\delta_{\bQ^\prime-\bQ,\bq_j}-\delta_{\bQ-\bQ^\prime,\bq_j}\right].
\ee
Moreover, the diamagnetic component of the current operator reads: 
\be
\mathcal{T}^{z,z}(\bq)=\sum_{\bk\in\text{MBZ}}\sum_{\bQ\bQ^\prime\in\mathcal{Q}_{1/2}}\sum_{\alpha,\beta}\,c^\dagger_{\bk,\bQ,\alpha}\,\tau^{z,z}_{\bQ\alpha,\bQ^\prime\beta}\,c^\dagga_{\bk+\bq,\bQ^\prime,\beta},
\ee
where
\be
\tau^{z,z}_{\bQ\alpha,\bQ^\prime\beta}=\frac{d^2_\perp}{\hbar^2}\,\sum_{j=1}^{3}T^j_{\alpha\beta}[\delta_{\bQ^\prime-\bQ,\bq_j}+\delta_{\bQ-\bQ^\prime,\bq_j}].
\ee
We observe that in this gauge the $\chi^{\parallel}_{\text{OMS}}$ is given by: 
\be
\label{OMS}
\chi^{\parallel}_{\text{OMS}}=-e^2\partial^2_q Q^{z,z}(q\,\mathbf{i},\omega=0)/2|_{q=0},
\ee  
where $\mathbf{i}$ is an in-plane vector of unit length, $Q^{z,z}(\bq,\omega=0)\equiv Q^{z,z}(\bq)$ is the static limit of the current response function
\bal
Q^{z,z}(\bq)&=-\frac{g_sg_v}{\pi}\int_{\text{MBZ}}\frac{d^2\bk}{4\pi^2}\,\int d\omega\,f(\omega)\\
&\Bigg\{\text{Im}\left[\text{Tr}\left( J^{p,z}G^R(\bk_+,\omega)\,J^{p,z}G^R(\bk_-,\omega)\right)\right]\\
&-\text{Im}\left[\text{Tr}\left(\tau^{z,z}\,G^{R}(\bk,\omega)\right)\right]\Bigg\},
\eal
with $\bk_\pm=\bk\pm\bq/2$, and $G^{R}(\bk,\omega)=\sum_n\ket{u_{n\bk}}\bra{u_{n\bk}}/(\omega-\ep_{n\bk}+i0^+)$ the single-particle Green's function.

\section{details on the evaluation of the intraband contribution $\chi^{\perp,\text{intra}}_{\text{OMS}}$}
\label{intraband_q_expansion}

In this section we provide analytic expressions for the inverse mass tensor $M^{-1}_{n\bk}$ \eqn{inverse_mass_tensor} and $\lambda^{(1)}_n(\bk)$ \eqn{lambda_1} in Eq. \eqn{OMS_perp_intra}.
To this aim we consider the perturbation: 
\bal
V_{\bQ\alpha,\bQ^\prime\beta}(\bq)&=H_{\bQ\alpha,\bQ^\prime\beta}(\bk+\bq)-H_{\bQ\alpha,\bQ^\prime\beta}(\bk)\\
&=\delta_{\bQ,\bQ^\prime}\hbar v_0 q_i\,\sigma^i_{\alpha\beta},
\eal
where $|\bq|/k_\theta\ll1$. 
For simplicity, we introduce the notation $V(\bq)=\hbar v_0\,q_i\,\Sigma^i$ where $\Sigma^i=\mathbf{1}\otimes\sigma^i$ is the tensor product between the identity in the lattice degree of freedom $\bQ$ and the Pauli matrix $\sigma^i$ ($i=x,y$) in the sublattice one. 

By employing non-degenerate perturbation theory on the eigenvalues $\ep_{n\bk}$ we find at second order in $q$:
\bal
&\ep_{n\bk\pm\bq}=\ep_{n\bk}\pm\hbar v_0\,q_i\bra{u_{n\bk}}\Sigma^i\ket{u_{n\bk}}\\
&+(\hbar\,v_0)^2\,q_i\,q_j\,\sum_{m\neq n}\frac{\bra{u_{n\bk}}\Sigma^i\ket{u_{m\bk}}\bra{u_{m\bk}}\Sigma^j\ket{u_{n\bk}}}{\ep_{n\bk}-\ep_{m\bk}},
\eal
so that $\partial_{k_i}\ep_{n\bk}=\hbar v_0 \bra{u_{n\bk}}\Sigma^i\ket{u_{n\bk}}$, and the components of the inverse mass tensor \eqn{inverse_mass_tensor} are 
\be
\partial^2_{k_i}\,\ep_{n\bk}=2(\hbar v_0)^2\sum_{m\neq n}\frac{|\Sigma^i_{nm}(\bk)|^2}{\ep_{n\bk}-\ep_{m\bk}},
\ee
and 
\be
\partial_{k_x}\partial_{k_y}\ep_{n\bk}=2(\hbar v_0)^2\sum_{m\neq n}\frac{\text{Re}\left[\Sigma^x_{nm}(\bk)\Sigma^y_{mn}(\bk)\right]}{\ep_{n\bk}-\ep_{m\bk}},
\ee
where $\Sigma^i_{nm}(\bk)=\bra{u_{n\bk}}\Sigma^i\ket{u_{m\bk}}$.
Moreover, we find:
 \be
 \ket{u^{(1)}_{n\bk}}=\frac{\hbar v_0}{2}\sum_{m\neq n}\frac{\Sigma^x_{mn}(\bk)}{\ep_n-\ep_m}\ket{u_{m\bk}},
 \ee
 and
 \bal
 \ket{u^{(2)}_{n\bk}}=\left(\frac{\hbar v_0}{2}\right)^2\sum_{m\neq n}&\Bigg\{\sum_{l\neq n}\frac{\Sigma^x_{ml}(\bk)\Sigma^x_{ln}(\bk)}{(\ep_{n\bk}-\ep_{m\bk})(\ep_{n\bk}-\ep_{l\bk})}\\
 &-\frac{\Sigma^x_{mn}(\bk)\Sigma^x_{nn}(\bk)}{(\ep_{n\bk}-\ep_{m\bk})^2}\Bigg\}\ket{u_{m\bk}}.
 \eal
By inserting the previous results in Eq. \eqn{lambda_1} we find: 
\bal
&\lambda^{(1)}_n(\bk)=-\frac{(\hbar v_0)^3}{4}\Sigma^y_{nn}(\bk)\sum_{m\neq n}\frac{|\Sigma^x_{mn}(\bk)|^2}{(\ep_{n\bk}-\ep_{j\bk})^2}\\
&-\frac{(\hbar v_0)^3}{4}\sum_{m\neq n}\frac{\Sigma^y_{nm}(\bk)\Sigma^x_{mn}(\bk)\Sigma^x_{nn}(\bk)+c.c.}{(\ep_{n\bk}-\ep_{m\bk})^2}\\
&+\frac{(\hbar v_0)^3}{4}\sum_{m,l\neq n}\frac{\Sigma^y_{nm}(\bk)\Sigma^x_{ml}(\bk)\Sigma^x_{ln}(\bk)+c.c.}{(\ep_{n\bk}-\ep_{m\bk})(\ep_{n\bk}-\ep_{l\bk})}\\
&-\frac{(\hbar v_0)^3}{4}\sum_{m,l\neq n}\frac{\Sigma^x_{nm}(\bk)\Sigma^y_{ml}(\bk)\Sigma^x_{ln}(\bk)}{(\ep_{n\bk}-\ep_{m\bk})(\ep_{n\bk}-\ep_{l\bk})},
\eal
and
\be
\lambda^{(2)}_n(\bk)=\frac{(\hbar v_0)^2}{2}\sum_{m\neq n}\frac{\Sigma^y_{nm}(\bk)\Sigma^{x}_{mn}(\bk)-c.c.}{\ep_{n\bk}-\ep_{m\bk}}.
\ee
As a consequence of  the $\mathcal{C}_{2z}T$ symmetry of the Hamiltonian $H_0$ \eqn{BM_TBG} the eigenstates are such that $u_{n\bk\bQ\alpha}=u^*_{n\bk\bQ\alpha^\prime}\sigma^{x}_{\alpha^\prime\alpha}$ where $\sigma^x$ exchanges $A$ and $B$ sublattices. 
By employing the previous symmetry we obtain that $\left[\Sigma^y_{nm}(\bk)\Sigma^{x}_{mn}(\bk)\right]^*=\Sigma^y_{nm}(\bk)\Sigma^{x}_{mn}(\bk)$ so that $\lambda^{(2)}_{n}(\bk)=0$.

\section{details on the evaluation of the interband contribution $\chi^{\perp,\text{inter}}_{\text{OMS}}$}
\label{interband_q_expansion}

In this section we give details on the evaluation of the interband contribution to $\chi^{\perp}_{\text{OMS}}$ \eqn{OMS_perp_inter}.
When a finite number of reciprocal lattice sites $\bQ$ are included the spectrum of TBG is composed by a finite number of eigenvalues $\ep_{n\bk}$ that we arrange in ascending order $n=1,\cdots,2N_L$ and $N_L$ is the number of lattice sites in reciprocal space.
We separate the contribution from the lowest positive energy band $\ep_{+\bk}$ ($n=N_L+1$ in the previous notation), that contains the Fermi energy, from the remaining bands:
\bal
\label{OMS_perp_inter_1}
\chi^{\perp,\text{inter}}_{\text{OMS}}=&g_sg_v\Bigg[\sum_{n=1}^{N_L}\sum_{n^\prime=N_L+2}^{2N_L} \chi^{(1)}_{n,n^\prime}\\
&+\sum_{n=1}^{N_L} \chi^{(1)}_{n,N_L+1}+\sum_{n^\prime=N_L+2}^{2N_L} \chi^{(2)}_{N_L+1,n^\prime}\Bigg],
\eal
where
\bal
\chi^{(1)}_{n,n^\prime}=&-\frac{e^2}{2}\frac{\partial^2}{\partial q^2}\sum_{\pm}\int_{\text{MBZ}} \frac{d^2\bk}{4\pi^2}[1-f(\ep_{n^\prime\bk})]\\
&\frac{|\bra{u_{n\bk\pm q\mathbf{x}}}J^y\ket{u_{n^\prime\bk}}|^2}{\ep_{n\bk\pm q\mathbf{x}}-\ep_{n^\prime\bk}}\Bigg|_{q=0},
\eal
and 
\bal
\chi^{(2)}_{N_L+1,n^\prime}=&-\frac{e^2}{2}\frac{\partial^2}{\partial q^2}\sum_{\pm}\int_{\text{MBZ}} \frac{d^2\bk}{4\pi^2}f(\ep_{N_L+1\bk})\\
&\frac{|\bra{u_{N_L+1\bk}}J^y\ket{u_{n^\prime\bk\pm q\mathbf{x}}}|^2}{\ep_{N_L+1\bk}-\ep_{n^\prime\bk\pm q\mathbf{x}}}\Bigg|_{q=0}.
\eal
In order to obtain $\chi^{(1)}_{n,n^\prime}$ and $\chi^{(2)}_{N_L+1,n^\prime}$ we have performed a Taylor expansion at second order in the wavevector $q$.
By following the same line of reasoning of Appendix \ref{intraband_q_expansion} we obtain the gauge-invariant expressions: 
\be
\chi^{(1)}_{n,n^\prime}=-\frac{e^2v^2_0}{2\pi^2}\int_{\text{MBZ}}d^2\bk\frac{1-f(\ep_{n^{\prime}\bk})}{\ep_{n\bk}-\ep_{n^\prime\bk}}\sum_{l=1}^{3}\Phi^{(l)}_{n,n^\prime}(\bk),
\ee
where 
\bal
&\Phi^{(1)}_{n,n^\prime}(\bk)=2\Re\Bigg[\Sigma^y_{n,n^\prime}(\bk)\sum_{m\neq n}\Sigma^{y}_{n^\prime,m}(\bk)\\
&\sum_{l\neq n}\frac{\Sigma^x_{m,l}(\bk)\Sigma^x_{l,n}(\bk)}{(\ep_{n\bk}-\ep_{m\bk})(\ep_{n\bk}-\ep_{l\bk})}-\Sigma^x_{n,n}(\bk)\Sigma^y_{n,n^\prime}(\bk)\\
&\sum_{m\neq n}\frac{\Sigma^y_{n^\prime,m}(\bk)\Sigma^x_{m,n}(\bk)}{\ep_{n\bk}-\ep_{m\bk}}\frac{2\ep_{n\bk}-\ep_{n^\prime\bk}-\ep_{m\bk}}{(\ep_{n\bk}-\ep_{n^\prime\bk})(\ep_{n\bk}-\ep_{m\bk})}\Bigg],
\eal
\be
\Phi^{(2)}_{n,n^\prime}(\bk)=\left|\sum_{m\neq n}\frac{\Sigma^x_{n,m}(\bk)\Sigma^y_{m,n^\prime}(\bk)}{\ep_{n\bk}-\ep_{m\bk}}\right|^2,
\ee
and
\bal
&\Phi^{(3)}_{n,n^\prime}(\bk)=\left|\Sigma^y_{n,n^\prime}(\bk)\right|^2\Bigg[\frac{\Sigma^x_{n,n}(\bk)^2}{(\ep_{n\bk}-\ep_{n^\prime\bk})^2}\\
&-\sum_{m\neq n}\frac{\left|\Sigma^x_{n,m}(\bk)\right|^2}{\ep_{n\bk}-\ep_{m\bk}}\frac{2\ep_{n\bk}-\ep_{n^\prime\bk}-\ep_{m\bk}}{(\ep_{n\bk}-\ep_{n^\prime\bk})(\ep_{n\bk}-\ep_{m\bk})}\Bigg].
\eal
Furthermore, by setting $\bar{n}=N_L+1$ we have 
\be
\chi^{(2)}_{\bar{n},n^\prime}=-\frac{e^2v^2_0}{2\pi^2}\int_{\text{MBZ}}d^2\bk\frac{f(\ep_{\bar{n}\bk})}{\ep_{\bar{n}\bk}-\ep_{n^\prime\bk}}\sum_{l=1}^{3}\Omega^{(l)}_{\bar{n},n^\prime}(\bk),
\ee
where
\bal
&\Omega^{(1)}_{\bar{n},n^\prime}(\bk)=2\Re\Bigg[\Sigma^y_{\bar{n},n^\prime}(\bk)\sum_{m\neq n^\prime}\Sigma^x_{n^\prime,m}(\bk)\\
&\sum_{l\neq n^\prime}\frac{\Sigma^x_{m,l}(\bk)\Sigma^y_{l,\bar{n}}(\bk)}{(\ep_{n^\prime\bk}-\ep_{m\bk})(\ep_{n^\prime}-\ep_{l\bk})}-\Sigma^y_{\bar{n},n^\prime}(\bk)\Sigma^x_{n^\prime,n^\prime}(\bk)\\
&\sum_{m\neq n^\prime}\frac{\Sigma^x_{n^\prime,m}(\bk)\Sigma^y_{m,\bar{n}}(\bk)}{(\ep_{n^\prime\bk}-\ep_{m\bk})^2}\frac{2\ep_{n^\prime\bk}-\ep_{m\bk}-\ep_{\bar{n}\bk}}{(\ep_{n^\prime\bk}-\ep_{\bar{n}\bk})}\Bigg],
\eal
\be
\Omega^{(2)}_{\bar{n},n^\prime}(\bk)=\left|\sum_{m\neq n^\prime}\frac{\Sigma^y_{\bar{n},m}(\bk)\Sigma^x_{m,n^\prime}(\bk)}{\ep_{n^\prime\bk}-\ep_{m\bk}}\right|^2,
\ee
and
\bal
&\Omega^{(3)}_{\bar{n},n^\prime}(\bk)=\left|\Sigma^y_{\bar{n},n^\prime}(\bk)\right|^2\Bigg[\frac{\Sigma^x_{n^\prime,n^\prime}(\bk)^2}{(\ep_{\bar{n}\bk}-\ep_{n^\prime\bk})^2}\\
&-\sum_{m\neq n^\prime}\frac{\left|\Sigma^x_{n^\prime,m}(\bk)\right|^2}{(\ep_{n^\prime\bk}-\ep_{m\bk})^2}\frac{2\ep_{n^\prime\bk}-\ep_{\bar{n}\bk}-\ep_{m\bk}}{(\ep_{n^\prime\bk}-\ep_{\bar{n}\bk})}\Bigg].
\eal

\section{finite size scaling for $\chi^{\perp,\text{inter}}_{\text{OMS}}$}
\label{finite_size_interlayer}

\begin{figure}
\begin{center}
\includegraphics[width=0.45\textwidth]{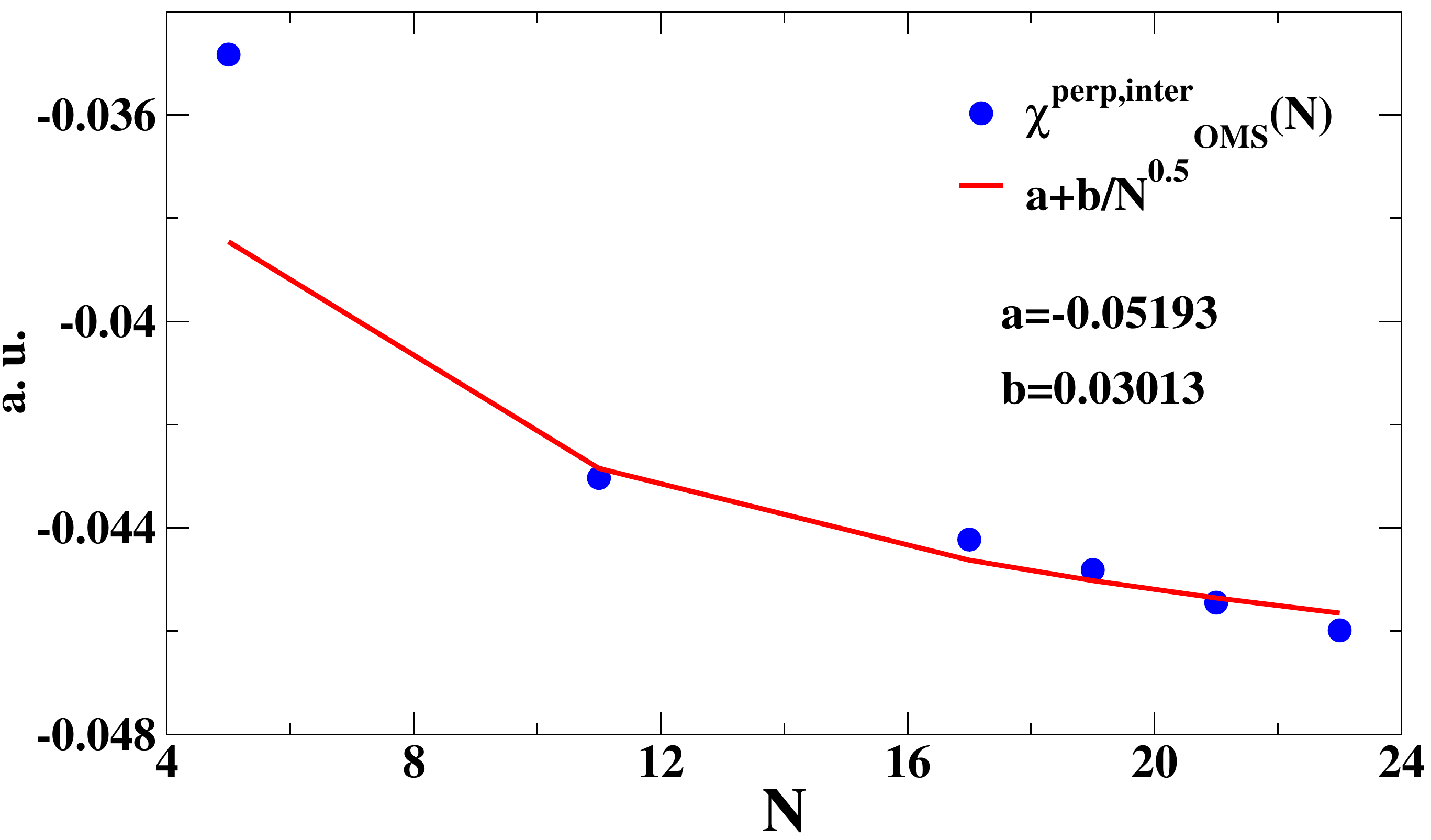}
\caption{ The blue dots are the numerical results for $\chi^{\perp,\text{inter}}_{\text{OMS}}(N,N)$ where $N^\prime$ has been taken large enough to converge the result in $N$ for $\theta=4^\circ$ $(\alpha=0.7)$, $r=0.7$ and $\mu/\hbar v_0 k_\theta=0.34$. The red line show the fit where the value of $a=\chi^{\perp,\text{inter}}_{\text{OMS}}=-0.05193$ (in arbitrary units) while the coefficient $b=0.03013$ which is not far from the one expected from Eq. \eqn{scaling_law}, i.e. $\sqrt{4\pi/3\sqrt{3}}/16\pi=0.03094$.}
\label{scaling_4_mu034}
\end{center}
\end{figure}

The evaluation of the interband contribution depends on the number of bands accounted in Eq. \eqn{OMS_perp_inter_1}. 
In order to achieve the convergence of $\chi^{\perp,\text{inter}}_{\text{OMS}}$ with the number of bands we introduce two integers $N_1$ and $N_2$, so that: 
\bal
\label{OMS_perp_inter_2}
&\chi^{\perp,\text{inter}}_{\text{OMS}}(N_1,N_2)=g_sg_v\Bigg[\sum_{n=N_L-N_1}^{N_L}\sum_{n^\prime=N_L+2}^{N_L+1+N_2}\chi^{(1)}_{n,n^\prime}\\
&+\sum_{n=N_L-N_2}^{N_L}\chi^{(1)}_{n,N_L+1}+\sum_{n^\prime=N_L+2}^{N_L+1+N_2} \chi^{(2)}_{N_L+1,n^\prime}\Bigg],
\eal
where $N_L\gg N_1, N_2$.
For a given value of $N_1$ we first converge $\chi^{\perp,\text{inter}}_{\text{OMS}}(N_1,N_2)$ in $N_2$. 
Then, to extract $\chi^{\perp,\text{inter}}_{\text{OMS}}$, i.e. the asymptotic value of ($N_1\gg1$) value of Eq. \eqn{OMS_perp_inter_2}, we perform a finite size scaling analysis as a function of $N_1$ by fitting the numerical result with $a+b/\sqrt{N_1}$.
The previous scaling law has been obtained by computing the interband contribution for two decoupled Dirac cones in the Moir\'e lattice when only $N_1$ bands below zero energy are accounted: 
\bal
\label{scaling_law}
\chi^{\perp,\text{inter}}_{\text{OMS}}(\alpha=0,\le N_1)=&-\frac{e^2v_0^2}{16\pi\mu}\\
&+\frac{e^2 v_0}{16\pi\hbar}\sqrt{\frac{4\pi}{3\sqrt{3}}}\frac{1}{\sqrt{N_1}}.
\eal
In Fig. \ref{scaling_4_mu034} we show the fit obtained for $\theta=4^\circ$ $(\alpha=0.7)$, $r=0.7$, the value of $b$ and the value of $\chi^{\perp,\text{inter}}_{\text{OMS}}$.

\section{regularization of $\chi^{\parallel,\text{inter}}_{\text{OMS}}$}
\label{reg_interlayer}

\begin{figure}
\begin{center}
\includegraphics[width=0.45\textwidth]{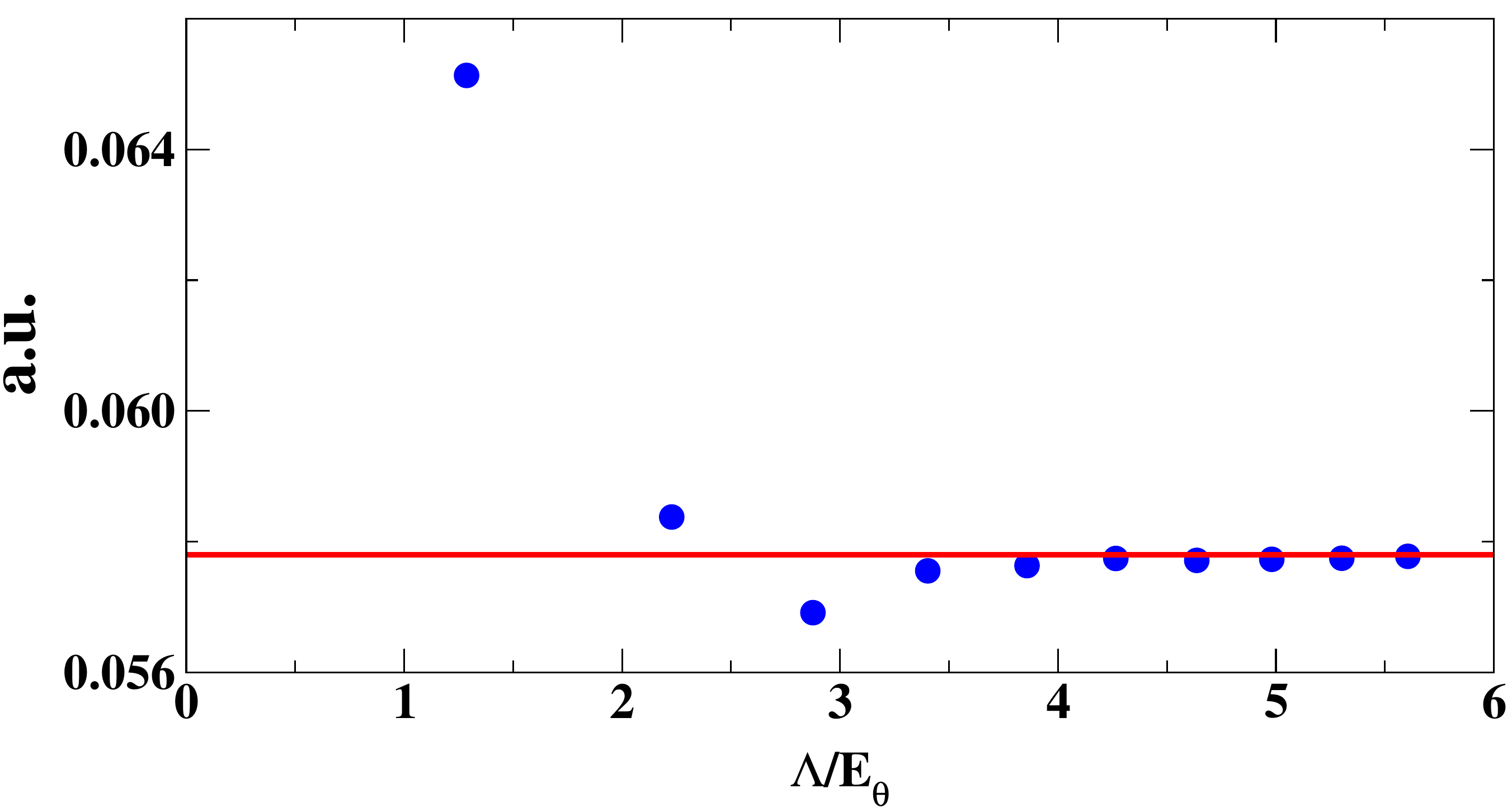}
\caption{The blue dots show the value of $\chi^{\parallel,\text{inter}}_{\text{OMS}}$ as a function of the UV cutoff $\Lambda/E_\theta$ where $E_\theta=\hbar v_0k_\theta$ for $r=0.2$, $\mu=0$ and $\theta=1.4^\circ$. The red solid line denotes the value of $\chi^{\parallel,\text{inter}}_{\text{OMS}}$ computed by employing Eq. \eqn{OMS}.}
\label{in_plane_UV_cutoff}
\end{center}
\end{figure}

In this Appendix we give details on the ultraviolet regularization scheme employed to compute $\chi^{\parallel,\text{inter}}_{\text{OMS}}$.
For a given value of the UV energy cutoff $\Lambda$ we compute the quantity $\chi^{\parallel,\text{inter}}_{\text{OMS}}=\overline{\chi}^{\parallel,\text{inter}}_{\text{OMS}}(\Lambda)-\overline{\chi}^{\parallel,\text{inter}}_{\text{OMS}}(\alpha=0,\Lambda)$, where we remind that $\overline{\chi}^{\parallel,\text{inter}}_{\text{OMS}}(\Lambda)$ is defined in Eq. \eqn{OMS_paral_inter_1} and $\overline{\chi}^{\parallel,\text{inter}}_{\text{OMS}}(\alpha=0,\Lambda)$ is computed by taking the eigenstates of two decoupled graphene layers.
Fig. \ref{in_plane_UV_cutoff} shows $\chi^{\parallel,\text{inter}}_{\text{OMS}}$ as a function of $\Lambda$ for $r=0.2$, $\theta=1.4^\circ$ and $\mu=0$.
The solid blue line shows the value of the in-plane orbital magnetic susceptibility \eqn{OMS} obtained in the gauge introduced in Appendix \ref{in_plane_Aperp_gauge}. 
As expected the data approach the solid blue line in the large $\Lambda$ regime which confirms the validity of the regularization scheme presented in Sec. \ref{OMS_in_plane}. 
 
\section{Relation between the counterflow conductivity $\sigma^{CF}$ and $\chi^{\parallel}_{\text{OMS}}$}
\label{counterflow_OMS}

In this Appendix we detail the relation between the counterflow conductivity $\sigma^{CF}(\omega)$ and the orbital magnetic response to an in-plane magnetic field.
The optical counterflow conductivity is given by: 
\be
\sigma^{CF}(\omega)=ie^2\,Q^{CF}(\omega)/(\omega+i0^+),
\ee
where 
\be
Q^{CF}(\omega)=\int dz\, z\int dz^\prime\,z^\prime\,Q^{y,y}(\bq=0,z,z^\prime,\omega).
\ee
Following Ref. \cite{Resta_2018} we write:
\be
\sigma^{CF}(\omega)=\left[\pi\,\delta(\omega)+i/\omega\right] D^{CF}+\sigma^{CF}_{\text{reg}}(\omega).
\ee 
In the previous Eq. $\sigma^{CF}_{\text{reg}}(\omega)$ is the regular part of $\sigma^{CF}(\omega)$, while $D^{CF}$ is the counterflow Drude weight
\be
D^{CF}=e^2\lim_{\omega\to0}\text{Re}\,Q^{CF}(\omega)
\ee
which gives the DC-conductivity $\sigma^{CF}=\tau D^{CF}$, where $\tau$ is a transport time introduced phenomenologically.  
Given the dipolar operator $\hat{\bm{\xi}}$ in Eq. \eqn{charge_dipolar}, which describes the current difference between top and bottom layers, we have: 
\bal
D^{CF}=&\frac{e^2g_vg_s}{4\pi^2}\sum_{n,n^\prime}^{n\neq n^\prime}\int_{\text{MBZ}}d^2\bk[f(\ep_{n\bk})-f(\ep_{n^\prime\bk})]\\
&\frac{ |\bra{u_{n\bk}}\xi^y\ket{u_{n^\prime\bk}}|^2}{\ep_{n\bk}-\ep_{n^\prime\bk}}=-\chi^{\parallel,\text{inter}}_{\text{OMS}},
\eal 
which coincides with the interband contribution to the orbital magnetic susceptibility $\chi^{\parallel}_{\text{OMS}}$. 
Therefore, at charge neutrality the counterflow Drude weight $D^{CF}$ is equal to $-\chi^{\parallel}_{\text{OMS}}$. 
Finally we observe that at finite chemical potential $D^{CF}$ differs from $\chi^{\parallel}_{\text{OMS}}$ by the Fermi surface contribution \eqn{OMS_paral_intra}, as a consequence of the dynamic limit ($\bq=0$ and $\omega\to0$) and static limit ($\omega=0$ and $\bq\to0$) taken to compute $D^{CF}$ and $\chi^{\parallel}_{\text{OMS}}$, respectively. 
 
 \section{Equation of motion of the Cavity Modes}
 \label{cavity_MODES}
 
By employing the Matsubara Green's function formalism~\cite{altland_simons_2010,Giuliani_book} we derive the equation of motion of the cavity modes propagators. Our aim is to find the instability condition towards the formation of an out of plane and in plane magnetic field. To start with we introduce the cavity modes Green's functions  
\bal
&\Pi^{XX}_{nm,\sigma}(\tau,\bq)=-\left\langle T_\tau\left(X_{\bq n\sigma}(\tau)\,X_{-\bq m\sigma}\right) \right\rangle,\\
&\Pi^{PX}_{nm,\sigma}(\tau,\bq)=-\left\langle T_\tau\left(P_{\bq n \sigma}(\tau)\,X_{-\bq m\sigma}\right) \right\rangle,
\eal
For the $\sigma=2$ (TE) modes we find: 
\bal
\label{Dyson_Pi_2}
\omega\,\Pi^{XX}_{nm,2}(\omega,\bq)&=i\omega_{\bq n}\,\Pi^{PX}_{nm,2}(\omega,\bq),\\
\omega\,\Pi^{PX}_{nm,2}(\omega,\bq)&=-i	\delta_{nm}-i\omega_{\bq n}\Pi^{XX}_{nm,2}(\omega,\bq)\\
-2e^2i\sum_{l}\alpha_{-\bq n1}&\,Q^L(\omega,\bq)\alpha_{\bq l1}\,\Pi^{XX}_{lm,1}(\omega,\bq)\\
-2e^2i\sum_{l}\alpha_{-\bq n2}&\,Q^T(\omega,\bq)\alpha_{\bq l2}\,\Pi^{XX}_{lm,2}(\omega,\bq),
\eal
where we have introduced the projection along the in-plane transverse and longitudinal directions of the current response tensor:
\bal
Q^T(\tau,\bq)&=\xi^i_{-\bq 2}\,Q^{i,j}(\tau,\bq)\,\xi^j_{\bq 2},\\
Q^L(\tau,\bq)&=\xi^i_{-\bq 1}\,Q^{i,j}(\tau,\bq)\,\xi^j_{\bq 1},
\eal
and $Q^{i,j}(\tau,\bq)=-\langle T_\tau\left(J^i(\bq,\tau)\,J^j(-\bq,\tau)\right)\rangle/S$. 
We are interested in magnetic instabilities of the systems that are signalled by the vanishing of the energy of the cavity modes. 
In the limit of $\omega\to0$ we have 
\be
\lim _{\omega\to0}Q^L(\omega,\bq)=0,
\ee
as a consequence of the the $f$-sum rule \cite{Giuliani_book}, so that only on the transverse response enters in the Dyson equation: 
\bal
&\sum_{l}\left[\frac{\omega^2-\omega^2_{n\bq}}{\omega_{n\bq}}\delta_{nl}-2e^2\alpha_{-\bq n 2}\,Q^T(\omega,\bq)\,\alpha_{\bq l 2}\right]\\
&\Pi^{XX}_{lm,2}(\omega,\bq)=\delta_{nm}.
\eal
For the $\sigma=1$ (TM) modes we find: 
\bal
\label{Dyson_Pi_1}
\omega\,\Pi^{XX}_{nm,1}(\omega,\bq)&=i\omega_{\bq n}\,\Pi^{PX}_{nm,1}(\omega,\bq),\\
\omega\,\Pi^{PX}_{nm,1}(\omega,\bq)&=-i	\delta_{nm}-i\omega_{\bq n}\Pi^{XX}_{nm,1}(\omega,\bq)\\
-2e^2i\sum_{l}\alpha_{-\bq n1}&\,Q^L(\omega,\bq)\alpha_{\bq l1}\,\Pi^{XX}_{lm,1}(\omega,\bq)\\
-2e^2i\sum_{l}\alpha^z_{-\bq n1}&\,Q^{z,z}(\omega,\bq)\alpha^z_{\bq l1}\,\Pi^{XX}_{lm,1}(\omega,\bq),
\eal
where we have introduced $Q^{z,z}(\tau,\bq)$ is 
\be
Q^{z,z}(\tau,\bq)=K^{z,z}(\tau,\bq)-\delta(\tau)\langle\mathcal{T}^{z,z}(0)\rangle/S,
\ee
and $K^{z,z}(\tau,\bq)=-\langle T_\tau\left(J^z(\bq,\tau)\,J^z(-\bq,\tau)\right)\rangle/S$. As a consequence of the $f$-sum rule the low-energy cavity photon properties are only affected by the transverse response:
\bal
&\sum_{l}\left[\frac{\omega^2-\omega^2_{n\bq}}{\omega_{n\bq}}\delta_{nl}-2e^2\alpha^z_{-\bq n 1}\,Q^{z,z}(\omega,\bq)\,\alpha^z_{\bq l 1}\right]\\
&\Pi^{XX}_{nm,1}(\omega,\bq)=\delta_{nm}.
\eal
Let us now look at the solution of the secular equations $\text{det}\left[\left(\bm{\Pi}^{XX}_{1/2}(\omega,\bq)\right)^{-1}\right]=0$.
By employing linear algebra theorems derived in Ref. \cite{Determinant_1960} we find: 
\bal
\text{det}\left[\left(\bm{\Pi}^{XX}_{1}(\omega,\bq)\right)^{-1}\right]&=\prod_n\,\frac{\omega^2-(\hbar\omega_{n\bq})^2}{\hbar\omega_{n\bq}}\\
\Bigg[1-2e^2\,Q^{z,z}(\omega,\bq)\sum_{n}&\frac{\alpha^z_{\bq n 1 }\,\alpha^z_{\bq n 1 }\,\hbar \omega_{n\bq}}{\omega^2-(\hbar \omega_{n\bq})^2}\Bigg]=0,
\eal
and
\bal
\text{det}\left[\left(\bm{\Pi}^{XX}_{2}(\omega,\bq)\right)^{-1}\right]&=\prod_n\,\frac{\omega^2-(\hbar\omega_{n\bq})^2}{\hbar\omega_{n\bq}}\\
\Bigg[1-2e^2\,Q^{\perp}(\omega,\bq)\sum_{n}&\,\frac{\alpha_{\bq n 2 }\,\alpha_{\bq n 2 }\,\hbar \omega_{n\bq}}{\omega^2-(\hbar \omega_{n\bq})^2}\Bigg]=0.
\eal
The in-plane and out of plane instabilities take place in the static limit $\omega=0$ when:
\be
\label{E_s1}
1+2e^2\,Q^{z,z}(\omega=0,\bq)\sum_{n}\,\frac{\alpha^z_{\bq n 1 }\,\alpha^z_{\bq n 1 }}{\hbar \omega_{n\bq}}\le 0,
\ee
and
\be
\label{E_s2}
1+2e^2\,Q^{\perp}(\omega=0,\bq)\sum_{n}\,\frac{\alpha_{\bq n 2 }\,\alpha_{\bq n 2 }}{\hbar \omega_{n\bq}}\le0.
\ee
The sum over the spectrum of cavity modes $n=0,1,\cdots$ is performed exactly. 
Eq. \eqn{E_s1} becomes 
\be
\label{E_s1_1}
\frac{\chi^\parallel_{\text{OMS}}(\bq)}{\chi_0(L_z)}\ge\frac{1}{1+\frac{qL_z}{4\pi}\coth\frac{qL_z}{2}+\frac{(q\,L_z)^2}{8\sinh(qL_z/2)^2}},
\ee
while for $\sigma=2$ we find
\be
\frac{\chi^\perp_{\text{OMS}}(\bq)}{\chi_0(L_z)}\ge\frac{1}{qL_z\,\tanh(q L_z/2)/2},
\ee
where $\chi_0(L_z)=L_z\,e^2 c/(4\pi \alpha_{EM}\hbar)$.
The latter expression recovers the instability criterion in Ref. \cite{Basko_PRL2019,Andolina_2020} for a spatially-modulated multimode cavity field.
In bilayer materials we introduce the instability criterion~\eqn{E_s1_1} for the photon condensation of the $\sigma=1$ cavity modes.

\bibliographystyle{apsrev}
\bibliography{mybiblio}
\end{document}